\documentclass[journal]{IEEEtran}   % IEEE Doublecolumn
\usepackage{array}
\usepackage{multirow}
\usepackage{diagbox}
\usepackage{amssymb}
\usepackage[caption=false,font=normalsize,labelfont=sf,textfont=sf]{subfig}
\usepackage{verbatim}
\usepackage{textcomp}
\usepackage{amsfonts}
\usepackage{amsmath}
\usepackage{graphicx}
\usepackage{cite}
\usepackage{balance}
\usepackage{epstopdf}
\usepackage{epsfig}
\usepackage{comment}
\usepackage{url}
\usepackage{xcolor}
\usepackage{fancyhdr}
\usepackage{mdwtab}
\usepackage{mathrsfs}
\usepackage{lipsum}
\usepackage{stfloats}
\usepackage[ruled,linesnumbered]{algorithm2e}
\SetAlCapSty{textbf}

\newlength{\figrowheight} % common image height for one-row figure* layouts

\newtheorem{theorem}{Theorem}

\newtheorem{lemma}{Lemma}

\newtheorem{corollary}{Corollary}

\newtheorem{remark}{Remark}  

\newtheorem{proposition}{Proposition}

\begin{document}
\bstctlcite{IEEEBibControl}
\title{Demand-Aware Beam Hopping Design for LEO Satellite Communications Using Movable Antennas}

\author{Jin Xie, Qimei Cui,~\IEEEmembership{Senior Member,~IEEE}, Xidong Mu,~\IEEEmembership{Member,~IEEE}, Xiaofeng Tao,~\IEEEmembership{Senior Member,~IEEE}, and Michail Matthaiou,~\IEEEmembership{Fellow,~IEEE} 

\thanks{Jin Xie, Qimei Cui and Xiaofeng Tao are with the National Engineering Research Center of Mobile Network Technologies, Beijing University of Posts and Telecommunications, Beijing 100876, China (e-mail: \{jin.xie, cuiqimei, taoxf\}@bupt.edu.cn). Qimei Cui and Xiaofeng Tao are also with the Department of Broadband Communication, Peng Cheng Laboratory, Shenzhen 518055, China.}
\thanks{Xidong Mu and Michail Matthaiou are with the Centre for Wireless Innovation (CWI), Queen's University Belfast, BT3 9DT Belfast, U.K. (e-mail: \{x.mu, m.matthaiou\}@qub.ac.uk). Michail Matthaiou is also with the Department of Electronic Engineering, Kyung Hee University, Yongin-si, Gyeonggi-do 17104, Korea.}
%\thanks{This work was carried out when J. Xie was visiting CWI funded by a CSC scholarship.}
%\thanks{Manuscript received April 19, 2021; revised August 16, 2021.}
}


% % column for its text to clear the IEEEpubid mark.

\maketitle

\begin{abstract}
Beam hopping (BH) enables low Earth orbit (LEO) satellites to allocate the limited beam resources among cells with uneven traffic demands. In line-of-sight-dominated satellite downlinks, however, fixed-position antenna (FPA) arrays struggle to suppress the inter-cell interference under the time-varying illumination requirements of BH. To address this challenge, this paper proposes a movable antenna (MA)-enabled demand-aware BH design for LEO satellite communications, which jointly optimizes the BH pattern, MA positions, and beamforming vectors to improve the supply-demand matching of each cell. A two-stage algorithm is developed to decompose the original problem into two subproblems. For the binary BH pattern optimization, a demand and aperture aware scheduling (DAAS) strategy is designed, in which a lower bound on the inter-cell channel correlation is derived from the MA movement region and cell directions to serve as a co-slot conflict penalty. By combining this penalty with the residual cell demands, the BH pattern is constructed slot by slot and then refined via cross-slot matching. For the joint beamforming vectors and MA positions optimization under the fixed BH pattern, a penalty dual decomposition-based algorithm is proposed, which decouples the two variables via auxiliary coefficients and solves the resulting augmented Lagrangian problem by block coordinate descent. Simulation results show that: 1) compared with the FPA scheme, the proposed design achieves smaller supply-demand gaps by adjusting the MA positions according to the illumination requirements of each time slot; 2) the adjusted MA positions enable the proposed design to focus the beams more precisely on the target cell centers, thereby reducing the interference leakage to other cells; and 3) the proposed DAAS strategy achieves more accurate demand matching than the round-robin and relaxation-based BH schemes.
\end{abstract}

\section{Introduction}
\IEEEPARstart{L}{ow} Earth orbit (LEO) satellite communications can provide access services to areas inadequately covered by terrestrial networks, serving as a key enabler of the ubiquitous connectivity envisioned for sixth-generation (6G) networks~\mbox{\cite{Cui2025AI6G,Luo2024LEOVLEOOverview,Wu2024SatelliteMIMOOverview}}. However, due to differences in population distribution and service types, the traffic demands of cells within the satellite coverage area typically exhibit significant non-uniformity \cite{Chen2026MultibeamHTSSurvey}. Moreover, the satellite onboard transmit power, processing capabilities, and the duration of service available for a given geographical region are limited \cite{Mahboob2024SatelliteNTNSurvey}. Therefore, dynamically allocating onboard resources in a demand-aware manner (within a service period) to match the communication capacity of each cell with its traffic demand remains a critical challenge in LEO satellite communications \cite{Chen2026AIBeamHoppingSurvey}. 

Beam hopping (BH) addresses this challenge by illuminating only a subset of cells in each time slot and adapting the illumination pattern across slots to the traffic demands~\cite{Lin2022DynamicBH}. Existing studies have jointly designed the BH pattern, time-slot allocation, and power allocation to match the offered capacity with these demands~\cite{Wang2022BH,Lin2024CoordinatedBH,Zheng2024TrafficAwareBH,Jia2025DynamicBH}. Regarding illumination scheduling, the authors of \cite{Lin2023MultiSatelliteBH} jointly considered multi-satellite load balancing and interference avoidance to improve the matching between offered capacity and traffic demand. Beyond illuminated-cell selection, the beam coverage radius was further optimized in \cite{Xu2023BHCoverage} to improve throughput and reduce packet loss. In \cite{Chen2024ProbabilisticBH}, a two-stage framework, based on beam activation probabilities, was developed to reduce energy consumption while satisfying the traffic demands. However, the data volume delivered to each cell also depends on the inter-beam interference among simultaneously illuminated cells, which tightly couples illumination scheduling with spatial-domain interference management.

Jointly designing the BH pattern and beamforming can suppress the interference among beams active in the same slot, thereby reducing the reliance of scheduling on the spatial separation of cells~\cite{Gupta2023ClusterHopping,Meng2025MAPPOBH,Wang2025CooperativeBH}. The authors of~\cite{Chen2023SelectiveBH} combined dynamic illumination with selective precoding, allowing adjacent beams to be simultaneously activated on demand. Under hybrid precoding, the BH pattern was designed by Viterbi and greedy algorithms to reduce the transmit power~\cite{Han2024HybridBH}, or jointly optimized with fully digital beamforming before constructing the hybrid beamformers to improve the sum rate~\cite{Wang2024JointBIP}. The number of users supported per unit power was further maximized in~\cite{Gong2026BH} by jointly optimizing beam selection, user scheduling, and precoding. However, all these studies employ fixed-position antenna (FPA) arrays, for which the array geometry and the cell directions jointly determine the inter-cell channel correlation under line-of-sight (LoS) propagation. Adjusting only the beamforming weights is insufficient to suppress the co-slot interference among highly correlated cells, which limits the ability of FPA arrays to underpin the time-varying illumination requirements of BH.

To overcome the fixed geometry of FPA arrays, movable antennas (MAs) and fluid antennas (FAs) have recently attracted significant attention~\cite{Wong2021FAS,New2025FASTutorial,Zhu2024MAOverview}. By adjusting the antenna positions within a confined region, MAs can obtain more favorable channel conditions, thereby introducing additional spatial degrees of freedom (DoFs)~\cite{Zhu2024MAOverview}, while prototypes based on motor-driven positioning platforms have validated the impact of position adjustment on spatial channels and communication performance~\cite{Dong2026MAPrototype}. Under LoS-dominated channels, an antenna displacement induces direction-dependent phase shifts across the array~\cite{Zhu2024MAModeling}, which allows MAs to reshape the spatial correlation among user channels~\cite{Zhu2026MATutorial}. By jointly optimizing the antenna positions and beamforming weights, MA arrays can enhance the gains toward multiple desired directions while suppressing interference in undesired ones~\cite{Ma2024MAMultiBeam,Zhu2023MANullSteering}, providing additional flexibility for simultaneously serving highly correlated cells.

Building on these advantages, MAs have been introduced into satellite communications to improve dynamic coverage and multi-user transmission performance. To cope with the coverage variations caused by satellite motion, the onboard antenna positions and beamforming weights were jointly optimized in~\cite{Zhu2025DynamicMA} to reduce the interference leakage outside the coverage area while satisfying the coverage gain requirements. In full-duplex satellite systems, the antenna positions and uplink/downlink power were further jointly optimized to mitigate self-interference and reduce the transmit power~\cite{Lin2025FullDuplexMA}. At ground stations, time-invariant antenna positions and time-varying beamforming were designed for dense LEO constellations in~\cite{Wang2026GroundStationMA}, and antenna positions, beamforming, and power splitting were jointly optimized for dual-hop satellite-terrestrial links in~\cite{Ali2026DualHopMA}. To mitigate uplink interference, the authors of~\cite{Han2026SatelliteCUMA} conceived an onboard compact ultra-massive FA array that suppresses inter-user interference by combining signals from multiple ports. Considering downlink multi-user access, the outage probability and ergodic data rate of FA-aided non-orthogonal multiple access transmission were derived in~\cite{Xie2026SatelliteFASNOMA} under spatially correlated shadowed-Ricean channels. 

\subsection{Motivation and Contributions}
Although the aforementioned studies have laid a foundation for BH resource scheduling and MA applications, how to coordinate BH scheduling with antenna position adjustment to improve demand matching across cells remains unexplored. Specifically, in LoS-dominated satellite channels, cells with similar directions tend to have highly correlated channels. However, conventional FPA arrays have limited beamforming DoFs, making it difficult to suppress the interference among these cells under the time-varying illumination requirements of BH. One way to avoid mutual interference is to schedule these cells in different slots, but this can limit how often each cell is illuminated and impair demand matching when their traffic demands are high~\cite{Chen2023SelectiveBH}. By adjusting the antenna positions according to the cells illuminated in each slot, MAs can dynamically reshape the inter-cell channel correlation and provide additional spatial DoFs for addressing this issue.

Exploiting this potential requires BH pattern generation to identify which cells can be spatially separated through MA position adjustment, whereas existing FPA-based designs do not take this factor into account. Moreover, these existing designs either evaluate numerous candidate illumination combinations at substantial computational cost~\cite{Wang2022BH} or rely on continuous relaxation, which reduces the complexity but may converge to local solutions with degraded performance~\cite{Wang2024JointBIP}. These issues are aggravated by MAs, whose positions further couple the discrete illumination decisions with the continuous transmission design. Therefore, a new low-complexity BH pattern generation algorithm is needed to account for the channel separation capability enabled by MA position adjustment.

Motivated by these considerations, we propose an MA-enabled demand-aware BH design for LEO satellite communications to improve the supply-demand matching of each cell within a service period. To this end, a two-stage algorithm is developed, where the BH pattern is first generated and the beamforming vectors and MA positions are then jointly optimized. The main contributions of this paper are summarized as follows:

\begin{enumerate}
  \item We consider a traffic supply-demand matching problem in an MA-enabled LEO satellite downlink system with BH. Given the transmit power budget and the maximum number of beams, we jointly optimize the BH pattern, MA positions, and beamforming vectors to minimize the sum of squared gaps between the cumulative delivered data volume and the traffic demand of each cell. We then introduce nonnegative variables for an equivalent reformulation and propose a two-stage optimization framework with low complexity.
  \item In the first stage, we propose a demand and aperture aware scheduling (DAAS) strategy to optimize the binary BH pattern. Based on the MA movement region and cell observation directions, DAAS computes a lower bound on the inter-cell channel correlation and uses it as a co-slot conflict penalty. Combining this bound with the residual cell demands, DAAS constructs an initial pattern slot by slot after reserving illumination opportunities for minimum service. The assignments are then refined via cross-slot matching to reduce the reference cumulative demand gaps.
  \item In the second stage, with the BH pattern fixed, a penalty dual decomposition (PDD)-based algorithm is employed to jointly optimize the beamforming vectors and the MA positions. We further introduce auxiliary variables to decouple the beamforming vectors and MA positions. The resulting augmented Lagrangian (AL) problem is solved by block coordinate descent (BCD), where the MA positions are updated via successive convex approximation (SCA) under the minimum spacing and movement speed constraints.
  \item Finally, simulation results are presented to validate the effectiveness of the MA-enabled demand-aware BH design. Compared with the FPA scheme, the proposed design achieves smaller supply-demand gaps by adjusting the MA positions according to the illumination requirements of each time slot. The adjusted MA positions enable the proposed design to focus the beams more precisely on the target cell centers, thereby reducing the interference leakage to other cells. Moreover, the proposed DAAS strategy achieves more accurate demand matching than the round-robin (RR) and relaxation-based BH schemes.
\end{enumerate}

\subsection{Organization and Notation}
The remainder of this paper is organized as follows: Section~II presents the system model and formulates the optimization problem. Section~III develops the two-stage algorithm and analyzes its convergence and computational complexity. Section~IV evaluates the supply-demand matching performance of the proposed scheme via simulations. Section~V concludes the paper. The proofs are provided in the Appendices.

\textit{Notation:} $(\cdot)^{\mathrm T}$, $(\cdot)^*$, and $(\cdot)^{\mathrm H}$ denote the transpose, conjugate, and conjugate transpose, respectively; $\Re\{\cdot\}$ and $\Im\{\cdot\}$ denote the real and imaginary parts; $\|\cdot\|_2$ denotes the Euclidean norm of a vector or the spectral norm of a matrix; $\|\cdot\|_1$ and $\|\cdot\|_\infty$ denote the $\ell_1$ and $\ell_\infty$ norms; $\|\cdot\|_F$ and $\|\cdot\|_{\max}$ denote the Frobenius norm and the maximum entry modulus of a matrix; $[\mathbf a]_{1:2}$ collects the first two entries of $\mathbf a$, and $\operatorname{diag}(\mathbf a)$ is the diagonal matrix with $\mathbf a$ on its diagonal. Vector inequalities are componentwise, and $\mathbf A\succ\mathbf 0$ means that $\mathbf A$ is Hermitian positive definite. We write $[a]_+=\max\{a,0\}$ for the positive part of $a$ and $[z]_a^b=\min\{b,\max\{a,z\}\}$ for the projection of $z$ onto $[a,b]$. The symbols $\lceil\cdot\rceil$ and $\lfloor\cdot\rfloor$ denote the ceiling and floor functions, respectively. Finally, $\mathbf1_{\{\cdot\}}$ denotes the indicator function.

\section{System Model}\label{System Model}
\begin{figure}[t!]
    \begin{center}
        \includegraphics[width=1\linewidth]{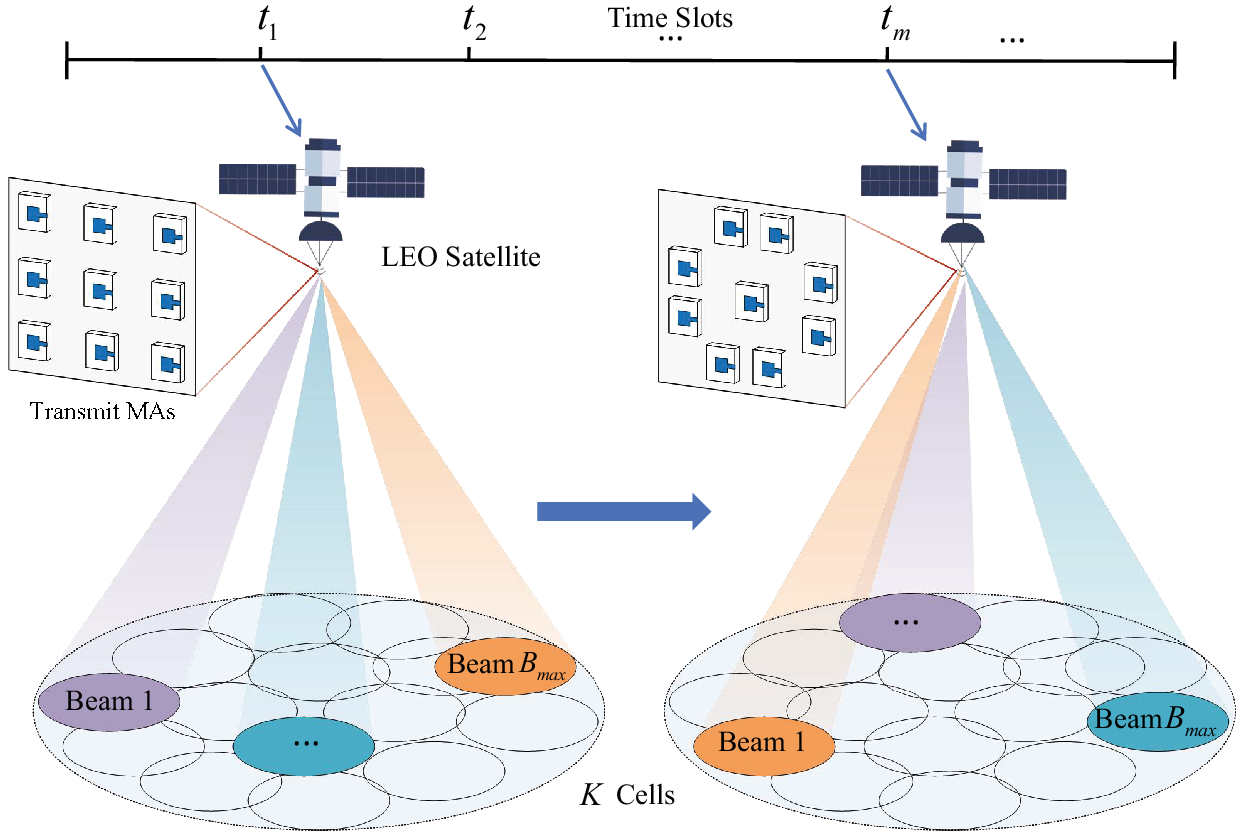}
        \caption{System model of the MA-enabled demand-aware BH design for LEO satellite communications.}
        \label{system_model_1}
    \end{center}
\end{figure}

We consider a LEO satellite providing downlink transmission over a region divided into $K$ cells, where the set of cells is denoted by $\mathcal K \triangleq \{1,\ldots,K\}$. Each cell is represented by a reference ground terminal located at the cell center, and $D_k$ denotes the cumulative traffic demand of cell $k$ over the service period, measured in bits. The LEO satellite is equipped with an $M$-element MA array and can simultaneously illuminate at most $B_{\max}$ cells, where $1\le B_{\max}\le K$. Each illuminated cell is assigned a dedicated transmit beam. The antenna index set is $\mathcal M\triangleq\{1,\ldots,M\}$. The satellite serves the considered region over a service period of duration $T_H$, which is divided into $N$ BH slots of equal duration $\Delta t = \frac{{{T_H}}}{N}$. Let $\mathcal N\triangleq  \left\{1,\ldots,N\right\}$ denote the slot set and $t_n=\left( {n - \frac{1}{2}} \right)\Delta t$ denote the midpoint of slot $n$. In each slot, the channel is approximated using the satellite position, attitude, and MA configuration at its midpoint. 
The binary BH indicator is denoted by $x_{k,n}\in\{0,1\}$, where $x_{k,n}=1$ indicates that cell $k$ is illuminated in slot $n$. The considered demand-aware design allows a high-demand cell to be illuminated repeatedly, whereas a cell whose demand has already been satisfied may remain non-illuminated. 

\subsection{Satellite Motion and Coordinate Transformation}
We establish a geocentric Cartesian coordinate system (GCCS) denoted by $O$-$XYZ$, where the $X$-$Y$ plane 
coincides with the equatorial plane and the $X$-axis points toward the ascending node of the satellite orbit. Let $H_{\mathrm s}$ and $\beta$ denote the orbital altitude and inclination angle, respectively. 
The orbital radius is then given by $R_{\mathrm s}=R_{\mathrm e}+H_{\mathrm s}$, where $R_{\mathrm e}$ is the radius of Earth.   
For a circular orbit, the angular velocity of the satellite is $\omega_{\mathrm s}=\sqrt{\mu/R_{\mathrm s}^{3}}$, where $\mu$ is Earth's gravitational parameter.
Accordingly, the orbital phase at slot $n$ can be expressed as
$\alpha(t_n)=\alpha_0+\omega_{\mathrm s}t_n$, where $\alpha_0$ is the initial orbital phase. 
The satellite position in the GCCS is therefore given by
$   {{\bf{r}}_{\rm{s}} \left( {{t_n}} \right)}= {R_{\rm{s}}}\left[ {\begin{array}{*{20}{c}}
{\cos \alpha \left( {{t_n}} \right)}&{\cos \beta \sin \alpha \left( {{t_n}} \right)}&{\sin \beta \sin \alpha \left( {{t_n}} \right)}
\end{array}} \right]^{\mathrm T}$. Cell $k$ on Earth's surface is represented by its angular coordinates $(\theta _k,\phi _k)$, where $\theta _k\in[-\pi/2,\pi/2]$ and $\phi _k\in(-\pi,\pi]$ denote the geocentric latitude and longitude, respectively. Its position vector in the GCCS is $  {{\bf{r}}_{k} }= {R_{\rm{e}}}\left[ {\begin{array}{*{20}{c}}
{\cos {\theta _k}\cos {\phi _k}}&{\cos {\theta _k}\sin {\phi _k}}&{\sin {\theta _k}}
\end{array}} \right]^{\mathrm T}$. 

To describe the orientation of the satellite-mounted MA array, a satellite-centric Cartesian coordinate system (SCCS) is introduced. 
Let ${{{{\bf{v}}_{\rm{s}}}\left( {{t_n}} \right)}}
=
\frac{\mathrm d\mathbf r_{\mathrm s}(t_n)}{\mathrm dt_n}$ denote the satellite velocity. The rotation matrix from the SCCS to the GCCS is $\mathbf U_n=[\mathbf e_x(t_n),\mathbf e_y(t_n),\mathbf e_z(t_n)]\in\mathbb R^{3\times3}$, where ${{\bf{e}}_x}\left( {{t_n}} \right) =\frac{{{{\bf{v}}_{\rm{s}}}\left( {{t_n}} \right)}}{{{{\left\| {{{\bf{v}}_{\rm{s}}}\left( {{t_n}} \right)} \right\|}_2}}}$ is aligned with the direction of motion, ${{\bf{e}}_z}\left( {{t_n}} \right) =  - \frac{{{{\bf{r}}_{\rm{s}}}\left( {{t_n}} \right)}}{{{R_{\rm{s}}}}}$ points toward the geocenter, and ${{\bf{e}}_y}\left( {{t_n}} \right) = {{\bf{e}}_z}\left( {{t_n}} \right) \times {{\bf{e}}_x}\left( {{t_n}} \right)$ is normal to the orbital plane. 

The wave vector from the satellite to cell $k$ in slot $n$, expressed in the SCCS, is given by \cite{Zhu2025DynamicMA}
\begin{equation}
\mathbf f_{k,n}
=
\frac{2\pi}{\lambda}
\mathbf U_n^{\mathrm T}
\frac{\mathbf d_{k,n}}
{\left\|\mathbf d_{k,n}\right\|_2},
\label{eq:local_wave_vector}
\end{equation} where $\lambda$ denotes the carrier wavelength and $\mathbf d_{k,n}=\mathbf r_{k}-\mathbf r_{\mathrm s} \left( {{t_n}} \right)$ is the displacement vector from the satellite to cell $k$.

\subsection{MA Positions and Channel Model}
The $M$ MAs can move within a prescribed two-dimensional region $\mathcal A$ on the satellite payload. Let $\mathbf {\bar z}_{m,n}=[{\mathbf z}_{m,n}^{\mathrm T},0]^{\mathrm T}\in\mathbb R^3$ denote the coordinates of the $m$-th MA element at slot $n$, where $\mathbf { z}_{m,n}=[x_m(t_n),y_m(t_n)]^{\mathrm T}\in\mathbb R^2$. 
The MA position matrix in slot $n$ collects the planar element positions as columns and can be written as 
\begin{equation}
    \mathbf Z_n
    =
    \begin{bmatrix}
        \mathbf { z}_{1,n} & \cdots & \mathbf { z}_{M,n}
    \end{bmatrix}
    \in\mathbb R^{2\times M}.
    \label{eq:apv_matrix}
\end{equation}

For the departure direction $\mathbf f_{k,n}$, the steering vector of the MA array is given by 
\begin{equation}
\mathbf v_{k,n}(\mathbf Z_n)=
\begin{bmatrix} {e^{j{\bf{f}}_{k,n}^{\rm{T}}{{{\bf{\bar z}}}_{1,n}}}} & \cdots & {e^{j{\bf{f}}_{k,n}^{\rm{T}}{{{\bf{\bar z}}}_{M,n}}}} 
\end{bmatrix} 
^{\rm{T}}.
\end{equation} 

Under a far-field narrowband LoS-dominated channel model with ideal Doppler compensation, the downlink channel from the satellite to cell $k$ in slot $n$ can be expressed as \cite{Zhu2025DynamicMA}
\begin{equation}
    \mathbf h_{k,n}(\mathbf Z_n)
    =
    \sqrt{\beta_{k,n}}
    e^{-j\frac{2\pi}{\lambda}\left\|\mathbf d_{k,n}\right\|_2}
    \mathbf v_{k,n}(\mathbf Z_n),
    \label{eq:los_channel}
\end{equation}
where the large-scale channel power gain is $  \beta_{k,n}=G_{\mathrm s}G_k\left(\frac{\lambda}{4\pi\left\|\mathbf d_{k,n}\right\|_2}\right)^2$, $G_{\mathrm s}$ is the gain of each satellite antenna and $G_k$ is the receive-antenna gain.

Assuming full frequency reuse among simultaneously illuminated cells, the signal received at the reference terminal of cell $k$ in slot $n$ is given by
\begin{align}
    y_{k,n}=\mathbf h_{k,n}^{\mathrm T}(\mathbf Z_n)\mathbf w_{k,n}s_k+\sum_{\ell\ne k}\mathbf h_{k,n}^{\mathrm T}(\mathbf Z_n)\mathbf w_{\ell,n}s_\ell+n_k,
\end{align} where $n_k\sim {\cal CN}(0,\sigma_k^2)$ denotes the additive white Gaussian noise, ${\bf w}_{k,n}\in {\mathbb C} ^{M \times 1} $ is the beamforming vector for cell $k$, and $s_k$ is the data symbol for cell $k$, normalized such that $\mathbb E[|s_k|^2]=1$.
The resulting signal-to-interference-plus-noise ratio (SINR) is 
\begin{align}
    \gamma_{k,n}=\frac{|\mathbf h_{k,n}^{\mathrm T}(\mathbf Z_n)\mathbf w_{k,n}|^2}{\sum_{\ell\ne k}|\mathbf h_{k,n}^{\mathrm T}(\mathbf Z_n)\mathbf w_{\ell,n}|^2+\sigma_k^2}.
\end{align}
The cumulative delivered data volume to cell $k$ over $T_H$ can be expressed as 
\begin{align}
    C_k=W\Delta t\sum_{n\in\mathcal N}\log_2(1+\gamma_{k,n}), 
\end{align} where $W$ denotes the bandwidth reused by the simultaneously illuminated cells.

\subsection{Problem Formulation}
Since the onboard power and beam resources may not fully accommodate the heterogeneous traffic demands, we minimize the sum of squared gaps between the cumulative delivered data volume and the traffic demand of each cell by jointly optimizing the BH pattern $\mathbf X\triangleq\{x_{k,n}\}_{k\in\mathcal K,n\in\mathcal N}$, beamforming vectors $\mathbf W\triangleq\{\mathbf w_{k,n}\}_{k\in\mathcal K,n\in\mathcal N}$, and MA positions $\mathbf Z\triangleq\{\mathbf Z_n\}_{n\in\mathcal N}$, which is formulated as
\begin{subequations}
\begin{align}
\mathcal P_0:\ \min\limits_{\mathbf X,\mathbf W,\mathbf Z}&\quad\sum_{k\in\mathcal K}(C_k-D_k)^2\\
\mathrm{s.t.}\quad
&{C_k} \ge C_k^{\min }, \forall k\in\mathcal K,
    \label{prob:P0_ck}\\
    &\sum_{k\in\mathcal K}x_{k,n}\leq B_{\max},\quad\forall n\in\mathcal N,
    \label{prob:P0_max_beams}\\
    &\sum_{k\in\mathcal K}\|{\bf w}_{k,n}\|_2^2
    \leq P_{\mathrm s},\quad \forall n\in\mathcal N,
    \label{prob:P0_power}\\
    &\left\| {{{\bf{w}}_{k,n}}} \right\|_2^2 \le {x_{k,n}}{P_{\mathrm s}},\quad \forall k,n, \label{prob:P0_power1}\\
     &x_{k,n}\in\{0,1\},\forall k\in\mathcal K,\ n\in\mathcal N,
    \label{prob:P0_binary}\\
    &{\bf z}_{m,n}\in\mathcal A,\forall m\in\mathcal M,\ n\in\mathcal N,
    \label{prob:P0_region}\\
    &\|{\bf z}_{m,n}-{\bf z}_{m',n}\|_2
    \geq d_{\min},\ \forall m\neq m',n,
    \label{prob:P0_spacing}\\
    &\|{\bf z}_{m,n}-{\bf z}_{m,n-1}\|_2\le v_{\max}\Delta t,\forall m,n\ge2,\label{prob:P0_speed}
\end{align}
\end{subequations}
where constraint \eqref{prob:P0_ck} ensures that the cumulative delivered data volume to each cell is at least $C_k^{\min}$, thereby satisfying its minimum quality-of-service (QoS) requirement; constraint \eqref{prob:P0_max_beams} ensures that the number of simultaneously illuminated cells does not exceed $B_{\max}$ in any time slot; constraint \eqref{prob:P0_power} guarantees that the satellite's transmit power does not exceed $P_{\mathrm s}$ in each time slot; constraint \eqref{prob:P0_power1} ensures zero beam power for each non-illuminated cell; constraint \eqref{prob:P0_binary} imposes binary illumination decisions; constraint \eqref{prob:P0_region} restricts each MA element to the movement region $\mathcal A$; constraint \eqref{prob:P0_spacing} ensures that the inter-element spacing is no smaller than the minimum allowable distance $d_{\min}$; and constraint \eqref{prob:P0_speed} limits the displacement of the MA elements between two consecutive time slots according to the maximum movement speed $v_{\max}$.

Problem $\mathcal P_0$ is a mixed-integer nonconvex optimization problem. The BH pattern determines the simultaneously illuminated cells, while the beamforming vectors and MA positions jointly control their desired signals and interference. Moreover, the movement-speed constraints couple the MA positions across slots, requiring their joint optimization over the entire service period. Another difficulty is that the sum of squared gaps is nonmonotonic in the cumulative delivered data volumes. Thus, directly replacing the achievable rates in the objective with concave lower bounds does not yield a valid objective upper bound. These difficulties make $\mathcal P_0$ challenging to solve directly.

\section{Two-Stage Optimization Algorithm}
\label{sec:algorithm}
To address the above challenges, we propose a two-stage algorithm that first constructs the BH pattern using DAAS, and then optimizes the beamforming vectors and MA positions via a PDD-based algorithm.

\subsection{Problem Reformulation}
To handle the nonmonotonic sum of squared gaps, we introduce nonnegative auxiliary variables and move the rate expressions into the constraints, where concave rate lower bounds can be applied.
Define $\kappa=W\Delta t$, $d_k=D_k/\kappa$, $c_k^{\min}=C_k^{\min}/\kappa$, and $\bar\delta_k=d_k-c_k^{\min}$, and assume $0\le C_k^{\min}\le D_k$ and $\sigma_k^2>0$ for all $k$. Using $\delta_k$ as a nonnegative upper bound on the traffic shortfall of cell $k$ normalized by $\kappa$, we reformulate $\mathcal P_0$ as 
\begin{subequations}\label{eq:deficit}
\begin{align}
\mathcal P_{\delta}:\quad
&\min\limits_{\mathbf X,\mathbf W,\mathbf Z,\boldsymbol\delta}
\sum_{k\in\mathcal K}\delta_k^2
\label{prob:Pdelta_obj}\\
\mathrm{s.t.}\quad
&\sum_{n\in\mathcal N}\log_2(1+\gamma_{k,n})
\ge d_k-\delta_k,\quad\forall k\in\mathcal K, 
\label{prob:Pdelta_service}\\
&0\le\delta_k\le\bar\delta_k,\quad\forall k\in\mathcal K,
\label{prob:Pdelta_bounds}\\
&\eqref{prob:P0_max_beams}\text{-}\eqref{prob:P0_speed}.\nonumber
\end{align}
\end{subequations}
Constraint \eqref{prob:Pdelta_service} gives $d_k-C_k/\kappa\le\delta_k$. Together with the upper bound $\delta_k\le\bar\delta_k$ in \eqref{prob:Pdelta_bounds}, this ensures $C_k\ge C_k^{\min}$. The following proposition establishes the relation between this formulation and $\mathcal P_0$.

\begin{proposition}\label{prop:reach}
If $\mathcal P_0$ admits an optimum, then $\mathcal P_0$ and $\mathcal P_\delta$ are equivalent in the following sense: their optimal values $v_0$ and $v_\delta$, respectively, satisfy $v_\delta=v_0/\kappa^2$. An optimum of $\mathcal P_0$ satisfying $C_k\le D_k$ for all $k$ can be recovered from any optimum of $\mathcal P_\delta$ by adjusting only the beam powers to attain $C_k^{\mathrm{tar}}=\min\{C_k,D_k\}$, without increasing any of them.
\end{proposition}
\begin{IEEEproof}
See Appendix~\ref{app:reach}.
\end{IEEEproof}

We therefore focus on solving $\mathcal P_\delta$ given in \eqref{eq:deficit}.

\subsection{DAAS-Based BH Pattern Generation}
In this subsection, we propose the DAAS strategy for BH pattern generation. It accounts for the impact of MA movement within the aperture on inter-cell channel correlation and combines this information with the remaining normalized traffic demands to select the illuminated cells in each slot. The strategy further refines the resulting cell-slot assignments across the service period using matching operations.

\subsubsection{Aperture-aware spatial conflict}
Let $\mathbf a_{k,n}=[\mathbf f_{k,n}]_{1:2}$ be the wave vector toward cell $k$ projected onto the MA plane, and define $\Delta\mathbf a_{ki,n}=\mathbf a_{i,n}-\mathbf a_{k,n}$. Under \eqref{eq:los_channel}, MA positions affect inter-cell channel correlation through the relative phases of the array responses. The normalized correlation is
\begin{equation}
 \chi_{ki,n}(\mathbf Z_n)
 =\left|\frac1M\sum_{m=1}^M
 e^{j\Delta\mathbf a_{ki,n}^{\mathrm T}\mathbf z_{m,n}}\right|^2.
 \label{eq:spatialcorr}
\end{equation}

For an arbitrary MA element position $\mathbf z=[x,y]^{\mathrm T}\in\mathcal A$, $\Delta\mathbf a_{ki,n}^{\mathrm T}\mathbf z$ is the position-dependent phase difference between the array responses toward cells $i$ and $k$. To score conflicts before MA optimization, define its aperture-permitted span as $\Omega_{ki,n}=\max_{\mathbf z\in\mathcal A}\Delta\mathbf a_{ki,n}^{\mathrm T}\mathbf z-\min_{\mathbf z\in\mathcal A}\Delta\mathbf a_{ki,n}^{\mathrm T}\mathbf z$. For a rectangular aperture with side lengths $L_x$ and $L_y$ along the $x$- and $y$-axes, $\Omega_{ki,n}=L_x|\Delta a_{ki,n,x}|+L_y|\Delta a_{ki,n,y}|$, where $\Delta a_{ki,n,x}$ and $\Delta a_{ki,n,y}$ are the $x$- and $y$-components of $\Delta\mathbf a_{ki,n}$, respectively. 
The following lemma lower-bounds this correlation over all aperture-feasible positions. 

\begin{lemma}\label{lem:aperture}
For all aperture-feasible positions, $\chi_{ki,n}(\mathbf Z_n)\ge\omega_{ki,n}$, where
\begin{equation}
 \omega_{ki,n}=\begin{cases}
 \cos^2(\Omega_{ki,n}/2),&\Omega_{ki,n}\le\pi,\\
 0,&\Omega_{ki,n}>\pi.
 \end{cases}
 \label{eq:conflict}
\end{equation}
\end{lemma}
\begin{IEEEproof}
When $\Omega_{ki,n}\le\pi$, rotating the center of the phasor arc onto the real axis makes every phasor's real part at least $\cos(\Omega_{ki,n}/2)\ge0$. The squared magnitude of their average therefore satisfies \eqref{eq:conflict}; otherwise, the zero lower bound applies.
\end{IEEEproof}

We then use $\omega_{ki,n}$ as a spatial-conflict penalty because a larger value indicates stronger channel correlation that MA movement cannot eliminate.

\subsubsection{Reference service and minimum-service quotas}
To evaluate cell illumination decisions, we use a reference array with fixed SCCS coordinates $\mathbf z_{m,n}^{\mathrm r}=\mathbf z_m^{\mathrm r}$ satisfying the aperture and inter-element spacing constraints. Its channels are evaluated using the satellite's position and attitude in each slot. Let $\widetilde{\mathcal S}_n\subseteq\mathcal K$ denote a candidate illuminated-cell set for slot $n$ and $\mathbf w_{k,n}^{\mathrm r}(\widetilde{\mathcal S}_n)$ denote the corresponding reference beamforming vector for cell $k$. For $|\widetilde{\mathcal S}_n|\le B_{\max}$, we adopt equal-power maximum-ratio transmission:
\begin{equation}
 \mathbf w_{k,n}^{\mathrm r}(\widetilde{\mathcal S}_n)=
 \begin{cases}
 \displaystyle\sqrt{p_{\mathrm r}}\,
 \frac{\mathbf h_{k,n}^*(\mathbf Z_n^{\mathrm r})}
 {\|\mathbf h_{k,n}(\mathbf Z_n^{\mathrm r})\|_2},&k\in\widetilde{\mathcal S}_n,\\[2mm]
 \mathbf0,&k\notin\widetilde{\mathcal S}_n,
 \end{cases}
 \label{eq:refbeam}
\end{equation}
where $p_{\mathrm r}=P_{\mathrm s}/B_{\max}$ is the fixed reference transmit power allocated to the beam serving each illuminated cell. Define $\chi_{ki,n}^{\mathrm r}=\chi_{ki,n}(\mathbf Z_n^{\mathrm r})$ and $\Gamma_{k,n}=p_{\mathrm r}M\beta_{k,n}/\sigma_k^2$. The normalized reference service for $k\in\widetilde{\mathcal S}_n$ is
\begin{equation}
 r_{k,n}^{\mathrm r}(\widetilde{\mathcal S}_n)
 =\log_2\left(1+\frac{\Gamma_{k,n}}
 {1+\Gamma_{k,n}\sum_{i\in\widetilde{\mathcal S}_n\setminus\{k\}}
 \chi_{ki,n}^{\mathrm r}}\right).
 \label{eq:refrate}
\end{equation}
Since each selected cell has at most $B_{\max}-1$ interferers, let $I_{k,n}^{\max}$ be the sum of the largest $B_{\max}-1$ entries of $\{\chi_{ki,n}^{\mathrm r}:i\ne k\}$, with $I_{k,n}^{\max}=0$ when $B_{\max}=1$. A slotwise service lower bound is
\begin{equation}
 \underline r_{k,n}
 =\log_2\left(1+\frac{\Gamma_{k,n}}
 {1+\Gamma_{k,n}I_{k,n}^{\max}}\right).
 \label{eq:guaranteedrates}
\end{equation}
The reference service satisfies $r_{k,n}^{\mathrm r}(\widetilde{\mathcal S}_n)\ge\underline r_{k,n}\ge\underline r_k$, where $\underline r_k=\min\limits_{n\in\mathcal N}\underline r_{k,n}$ is the uniform lower bound across slots.

At the start of slot $n$, the accumulated normalized reference data volume $\widehat s_{k,n}$ determines the remaining normalized traffic demand $q_{k,n}^{\mathrm D}=[d_k-\widehat s_{k,n}]_+$ and the remaining normalized QoS requirement $q_{k,n}^{\min}=[c_k^{\min}-\widehat s_{k,n}]_+$. We assume that the QoS requirements are compatible with the reference service, i.e., $\underline r_k>0$ whenever $c_k^{\min}>0$. The remaining illumination quota is then given by
\begin{equation}
 \ell_{k,n}=
 \begin{cases}
 \left\lceil q_{k,n}^{\min}/\underline r_k\right\rceil,
 &q_{k,n}^{\min}>0,\\
 0,&q_{k,n}^{\min}=0.
 \end{cases}
 \label{eq:residualdemands}
\end{equation}
We further assume that the initial quotas satisfy $\ell_{k,1}\le N$ and $\sum_k\ell_{k,1}\le NB_{\max}$, i.e., the QoS requirements can be accommodated by the available slots and beams.

\subsubsection{Quota-protected cell selection}
With $\mu_{k,n}=\min\{q_{k,n}^{\mathrm D},\underline r_{k,n}\}$, a lower bound on the reduction in $(q_{k,n}^{\mathrm D})^2$ is $g_{k,n}=(q_{k,n}^{\mathrm D})^2-(q_{k,n}^{\mathrm D}-\mu_{k,n})^2=2q_{k,n}^{\mathrm D}\mu_{k,n}-\mu_{k,n}^2$, and the slotwise surrogate is
\begin{equation}
 J_n(\widetilde{\mathcal S}_n)
 =\sum_{k\in\widetilde{\mathcal S}_n}g_{k,n}
 -\lambda_n^{\mathrm{sp}}
 \sum_{\substack{k<i\\k,i\in\widetilde{\mathcal S}_n}}\omega_{ki,n}.
 \label{eq:patternutility}
\end{equation}
The marginal gain of adding cell $k\notin\widetilde{\mathcal S}_n$ is
\begin{equation}
 \Delta_{k,n}(\widetilde{\mathcal S}_n)
 =g_{k,n}-\lambda_n^{\mathrm{sp}}
 \sum_{i\in\widetilde{\mathcal S}_n}\omega_{ki,n},
 \label{eq:marginal}
\end{equation}
where $\lambda_n^{\mathrm{sp}}=\eta_{\mathrm{sp}}\max\limits_k g_{k,n}/(B_{\max}-1)$ for $B_{\max}>1$ and $\lambda_n^{\mathrm{sp}}=0$ otherwise, with $\eta_{\mathrm{sp}}\ge0$. Adding cell $j$ updates each remaining score as $\Delta_{k,n}\gets\Delta_{k,n}-\lambda_n^{\mathrm{sp}}\omega_{kj,n}$.

Since the reference service is at least $\underline r_{k,n}$, $J_n(\widetilde{\mathcal S}_n)$ lower-bounds the one-slot reduction in $\sum_k(q_{k,n}^{\mathrm D})^2$. With $R_n=N-n+1$ remaining slots, define $\mathcal U_n=\{k:\ell_{k,n}>0\}$ and $\mathcal F_n=\{k:\ell_{k,n}=R_n\}$, and set
\begin{equation}
 L_n^{\mathrm{req}}
 =\max\left\{|\mathcal F_n|,
 \left[\sum_k\ell_{k,n}-B_{\max}(R_n-1)\right]_+\right\}.
\label{eq:requiredcount}
\end{equation}
Cells in $\mathcal F_n$ cannot be postponed, and $[\sum_k\ell_{k,n}-B_{\max}(R_n-1)]_+$ counts quotas exceeding future capacity. Thus, the slot must serve at least $L_n^{\mathrm{req}}$ positive-quota cells:
\begin{subequations}\label{eq:patternsurrogate}
\begin{align}
 \max\limits_{\widetilde{\mathcal S}_n\subseteq\mathcal K}\quad&J_n(\widetilde{\mathcal S}_n)
 \label{eq:patternsurrogate_obj}\\
 \mathrm{s.t.}\quad&\mathcal F_n\subseteq\widetilde{\mathcal S}_n,
 \label{eq:patternsurrogate_mandatory}\\
 &|\widetilde{\mathcal S}_n|\le B_{\max},
 \label{eq:patternsurrogate_capacity}\\
 &|\widetilde{\mathcal S}_n\cap\mathcal U_n|\ge L_n^{\mathrm{req}}.
 \label{eq:patternsurrogate_quota}
\end{align}
\end{subequations}

For each slot, DAAS constructs a feasible set for \eqref{eq:patternsurrogate} using the marginal gains in \eqref{eq:marginal}, which measure the change in $J_n$ from adding a cell. Starting from $\widetilde{\mathcal S}_n=\mathcal F_n$, it adds the cell in $\mathcal U_n\setminus\widetilde{\mathcal S}_n$ with the largest marginal gain until \eqref{eq:patternsurrogate_quota} holds, even if the gain is nonpositive. It then repeatedly adds the unselected cell with the largest positive marginal gain until $B_{\max}$ cells have been selected or no positive gain remains. Ties are broken by cell index. The resulting set is denoted by $\mathcal S_n$, with $x_{k,n}=\mathbf1_{\{k\in\mathcal S_n\}}$, and the cumulative reference service is updated as $\widehat s_{k,n+1}=\widehat s_{k,n}+r_{k,n}^{\mathrm r}(\mathcal S_n)$, where non-illuminated cells receive zero service.

% If the reference array is geometrically feasible, $\underline r_k>0$ whenever $c_k^{\min}>0$, and the initial quota conditions hold, Algorithm~\ref{alg:DAAS} produces a feasible reference implementation $(\mathbf X,\mathbf W^{\mathrm r},\mathbf Z^{\mathrm r})$ of $\mathcal P_0$. Each selected positive-quota cell receives at least $\underline r_k$, reducing its quota by at least one. Including $\mathcal F_n$ and serving at least $L_n^{\mathrm{req}}$ such cells preserve $\ell_{k,n}\le R_n$ and $\sum_k\ell_{k,n}\le B_{\max}R_n$, which imply $|\mathcal F_n|\le L_n^{\mathrm{req}}\le\min\{B_{\max},|\mathcal U_n|\}$ and ensure feasible selections. Consequently, all terminal quotas vanish and $\widehat s_{k,N+1}\ge c_k^{\min}$; the remaining constraints hold by the reference construction.

\subsubsection{Matching-based BH pattern refinement}
\label{sec:matching}
Under the above assumptions, the slotwise selections yield a feasible initial matching $\mathbf X^{(0)}$. DAAS refines this matching across all slots by minimizing the reference demand deficit
\begin{equation}
\Phi(\mathbf X)=\sum_{k\in\mathcal K}
\left[d_k-s_k^{\mathrm r}(\mathbf X)\right]_+^2,
\label{eq:matching_score}
\end{equation}
where $s_k^{\mathrm r}(\mathbf X)=\sum_{n:x_{k,n}=1}r_{k,n}^{\mathrm r}(\mathcal S_n)$ is the cumulative normalized reference service. To preserve the initial pattern conflict level, we constrain the aggregate conflict over pairs of simultaneously illuminated cells, $C_\omega(\mathbf X)=\sum_{n\in\mathcal N}\sum_{k<i}x_{k,n}x_{i,n}\omega_{ki,n}$, by its initial value $C_0=C_\omega(\mathbf X^{(0)})$, which is fixed throughout refinement.

A replacement moves the illumination of cell $i$ in slot $n$ to a non-illuminated cell-slot pair $(j,m)$, where $m$ may equal $n$. A swap exchanges the slots of two illuminated cells, i.e., moves $(i,n)$ and $(j,m)$ to $(i,m)$ and $(j,n)$ if both are non-illuminated. Both types of moves preserve the total number of illuminations. Admissible moves satisfy the per-slot illumination limit, $s_k^{\mathrm r}(\mathbf X)\ge c_k^{\min}$ for all cells, and $C_\omega(\mathbf X)\le C_0$, preserving reference feasibility and the initial conflict bound.

DAAS accepts an admissible move only if it reduces $\Phi$ by at least $\epsilon_{\mathrm{match}}$, which is set to $10^{-4}\max\{1,\Phi(\mathbf X^{(0)})\}$. It repeatedly applies the best replacement until none qualifies, then accepts the best qualifying swap and returns to replacements. The search stops when neither type of move qualifies. Algorithm~\ref{alg:DAAS} summarizes the complete DAAS procedure and returns the final BH pattern $\mathbf X$, which is fixed for the subsequent PDD optimization of the beamforming vectors and MA positions.

\begin{algorithm}[t]
\DontPrintSemicolon
\SetAlgoNoLine
\SetKwInOut{KwIn}{Input}
\SetKwInOut{KwOut}{Output}
\SetNlSty{textnormal}{}{:}
\caption{DAAS-Based BH Pattern Generation}
\label{alg:DAAS}
\KwIn{$\mathcal K,\allowbreak\mathcal N,\allowbreak M,\allowbreak\mathcal A,\allowbreak\mathbf Z^{\mathrm r},\allowbreak P_{\mathrm s},\allowbreak B_{\max},\allowbreak\eta_{\mathrm{sp}},\allowbreak\epsilon_{\mathrm{match}};\allowbreak\{\mathbf a_{k,n},\beta_{k,n}\}_{k,n};\allowbreak\{d_k,c_k^{\min},\sigma_k^2\}_k$.}
Compute $\chi_{ki,n}^{\mathrm r}$, $\Omega_{ki,n}$, and $\omega_{ki,n}$ by \eqref{eq:spatialcorr}-\eqref{eq:conflict}\;
Set $p_{\mathrm r}=P_{\mathrm s}/B_{\max}$ and $\Gamma_{k,n}=p_{\mathrm r}M\beta_{k,n}/\sigma_k^2$; form $I_{k,n}^{\max}$, compute $\underline r_{k,n}$ by \eqref{eq:guaranteedrates}, and set $\underline r_k=\min\limits_{n\in\mathcal N}\underline r_{k,n}$\;
Initialize $\widehat s_{k,1}=0$, $q_{k,1}^{\min}=c_k^{\min}$, and $\ell_{k,1}$ by \eqref{eq:residualdemands}\;
\For{$n=1,\ldots,N$}{
 Update $q_{k,n}^{\mathrm D}$ and $q_{k,n}^{\min}$; compute $\ell_{k,n}$ by \eqref{eq:residualdemands}\;
 Compute $\mu_{k,n}$, $g_{k,n}$, and $\lambda_n^{\mathrm{sp}}$; form $R_n$, $\mathcal U_n$, $\mathcal F_n$, and $L_n^{\mathrm{req}}$\;
 Set $\widetilde{\mathcal S}_n\gets\mathcal F_n$ and initialize $\Delta_{k,n}$ by \eqref{eq:marginal}\;
 Add cells of $\mathcal U_n$ with the largest $\Delta_{k,n}$ until \eqref{eq:patternsurrogate_quota} holds; then add cells with the largest positive $\Delta_{k,n}$ while $|\widetilde{\mathcal S}_n|<B_{\max}$, updating $\Delta_{k,n}$ after each addition\;
 Set $\mathcal S_n\gets\widetilde{\mathcal S}_n$ and $x_{k,n}=\mathbf1_{\{k\in\mathcal S_n\}}$; update $\widehat s_{k,n+1}$ by \eqref{eq:refrate}\;
}
Set $\mathbf X^{(0)}\gets\mathbf X$ and $C_0=C_\omega(\mathbf X^{(0)})$\;
\While{an admissible replacement or swap decreases $\Phi$ by at least $\epsilon_{\mathrm{match}}$}{
 Apply the best such replacement if one exists, and otherwise the best such swap; update the affected slots\;
}
\KwOut{$\mathbf X$.}
\end{algorithm}

\subsection{PDD-Based Optimization of $\mathbf W$ and $\mathbf Z$}
To separate the rate constraints from the coupling between beamforming and MA positions, we introduce an auxiliary matrix of effective signal coefficients, which can be expressed as 
\begin{equation}
 \mathbf Q_n=\mathbf H_n(\mathbf Z_n)\mathbf W_n\in\mathbb C^{b_n\times b_n},
 \label{eq:qcons}
\end{equation}
where $b_n=|\mathcal S_n|$, $\mathbf H_n=[\mathbf h_{k,n}^{\mathrm T}(\mathbf Z_n)/\sigma_k]_{k\in\mathcal S_n}\in\mathbb C^{b_n\times M}$ and $\mathbf W_n=[\mathbf w_{k,n}]_{k\in\mathcal S_n}\in\mathbb C^{M\times b_n}$. 
With $q_{ki,n}=[\mathbf Q_n]_{k,i}$ and $\mathbf q_{k,n}=[q_{ki,n}]_{i\in\mathcal S_n}^{\mathrm T}$, the normalized data volume (in bit/s/Hz) for illuminated cell $k$ in slot $n$ can be expressed as
\begin{equation}
 r_{k,n}(\mathbf q_{k,n})
 =\log_2\left(1+\frac{|q_{kk,n}|^2}
 {1+\sum_{i\in\mathcal S_n\setminus\{k\}}|q_{ki,n}|^2}\right).
 \label{eq:qrate}
\end{equation}
When \eqref{eq:qcons} holds, the data rate is $W r_{k,n}$ bit/s and the slot delivers $W\Delta t\,r_{k,n}=\kappa r_{k,n}$~bits. Thus, $C_k/\kappa$ equals the sum of $r_{k,n}$ over the slots in which cell $k$ is illuminated.

With $\mathbf X$ fixed, the beamforming vectors of non-illuminated cells are set to zero. Hence, \eqref{prob:P0_power1} is automatically satisfied, only the beams of cells in $\mathcal S_n$ generate interference in \eqref{eq:qrate}, and \eqref{prob:P0_power} reduces to a power constraint on $\mathbf W_n$.

The proposed PDD method incorporates the consistency constraints \eqref{eq:qcons} into an AL~\cite{Shi2020PDD1}. For a fixed penalty parameter $\rho>0$ and multiplier matrices $\{\mathbf\Lambda_n\}$, the AL optimization problem can be expressed as
\begin{subequations}\label{eq:ALproblem}
\begin{align}
 \min\limits_{\mathcal V}\quad
 &\mathcal L_\rho(\mathcal V)=\sum_k\delta_k^2+\frac1{2\rho}\sum_n
 \|\mathbf R_n+\rho\mathbf\Lambda_n\|_F^2
 \label{eq:AL}\\
 \mathrm{s.t.}\quad
 &\sum_{n\in\mathcal N_k}r_{k,n}(\mathbf q_{k,n})
 \ge d_k-\delta_k,\quad\forall k,
 \label{eq:fixedproblem_service}\\
 &0\le\delta_k\le\bar\delta_k,\quad\forall k,
 \label{eq:fixedproblem_bounds}\\
 &\|\mathbf W_n\|_F^2\le P_{\mathrm s},\quad\forall n,
 \label{eq:fixedproblem_power}\\
 &\eqref{prob:P0_region}\text{-}\eqref{prob:P0_speed},
 \label{eq:AL_geometry}
\end{align}
\end{subequations}
where $\mathcal V=\{\mathbf Q,\boldsymbol\delta,\mathbf W,\mathbf Z\}$, $\mathcal N_k=\{n:x_{k,n}=1\}$ is the set of slots serving cell $k$, $\mathbf R_n=\mathbf Q_n-\mathbf H_n(\mathbf Z_n)\mathbf W_n$ is the slotwise consistency residual, and $\mathbf\Lambda_n\in\mathbb C^{b_n\times b_n}$ contains the associated Lagrange multipliers. 

At PDD outer iteration $t$, fix $\rho=\rho_t$ and $\mathbf\Lambda_n=\mathbf\Lambda_n^t$. The outer updates of $\rho$ and $\{\mathbf\Lambda_n\}$ are designed to reduce the consistency residuals. An inner BCD loop, indexed by $j$, approximately solves \eqref{eq:ALproblem} by successively updating $(\mathbf Q,\boldsymbol\delta)$, $\mathbf W$, and $\mathbf Z$, as detailed below.

\subsubsection{Subproblem with respect to $\{\mathbf Q,\boldsymbol\delta\}$}
With $\mathbf W$ and $\mathbf Z$ fixed, the quadratic penalty term in \eqref{eq:AL} penalizes deviations of $\mathbf Q_n$ from the fixed matrix $\mathbf T_n=\mathbf H_n(\mathbf Z_n)\mathbf W_n-\rho\mathbf\Lambda_n$. The resulting $(\mathbf Q,\boldsymbol\delta)$ subproblem can be written as
\begin{align}
 \min\limits_{\mathbf Q,\boldsymbol\delta}\quad
 &\sum_k\delta_k^2
 +\frac1{2\rho}\sum_n\|\mathbf Q_n-\mathbf T_n\|_F^2
 \label{eq:qoriginal}\\ 
 \mathrm{s.t.}\quad
 &\eqref{eq:fixedproblem_service},\quad\eqref{eq:fixedproblem_bounds}.\nonumber
\end{align}

Since $r_{k,n}$ in \eqref{eq:fixedproblem_service} is nonconcave in $\mathbf q_{k,n}$, we construct a concave lower bound using the weighted minimum mean-square error (WMMSE) representation~\cite{Shi2011WMMSE}. For a complex receive coefficient $u_{k,n}$, the mean-square error can be expressed as
\begin{equation}
\begin{aligned}
 e_{k,n}(u_{k,n},\mathbf q_{k,n})
 ={}&|u_{k,n}|^2\left(1+\sum_{i\in\mathcal S_n}|q_{ki,n}|^2\right)\\
 &-2\Re\{u_{k,n}^*q_{kk,n}\}+1.
\end{aligned}
\label{eq:mse}
\end{equation}
With a positive weight $v_{k,n}$, the WMMSE identity gives
\begin{equation}
 r_{k,n}(\mathbf q_{k,n})
 =\max\limits_{\substack{u_{k,n}\in\mathbb C\\v_{k,n}>0}}
 \frac{\ln v_{k,n}-v_{k,n}e_{k,n}+1}{\ln2}.
 \label{eq:wmmse}
\end{equation}
At each BCD iteration, the receive coefficient for the current feasible signal vector $\mathbf q_{k,n}^{\mathrm{cur}}$ can be updated as
\begin{equation}
 u_{k,n}
 =\frac{q_{kk,n}^{\mathrm{cur}}}
 {1+\sum_{i\in\mathcal S_n}|q_{ki,n}^{\mathrm{cur}}|^2}.
 \label{eq:uvupdate}
\end{equation}
The corresponding weight is $v_{k,n}=1/e_{k,n}(u_{k,n},\mathbf q_{k,n}^{\mathrm{cur}})$. The lower-bound coefficients are $a_{k,n}=v_{k,n}|u_{k,n}|^2/\ln2$ and $b_{k,n}=v_{k,n}u_{k,n}/\ln2$, with the constant term given by
\begin{equation}
 c_{k,n}=
 \frac{\ln v_{k,n}+1-v_{k,n}(1+|u_{k,n}|^2)}{\ln2}.
 \label{eq:ratecoefficients}
\end{equation}
The resulting lower bound on $r_{k,n}$ can be expressed as
\begin{equation}
\begin{aligned}
 \widetilde r_{k,n}(\mathbf q_{k,n})
 ={}c_{k,n}-a_{k,n}\sum_{i\in\mathcal S_n}|q_{ki,n}|^2+2\Re\{b_{k,n}^*q_{kk,n}\}.
\end{aligned}
\label{eq:lb}
\end{equation}
Since $a_{k,n}\ge0$, the bound is concave. It matches the value and gradient of $r_{k,n}$ at the current point, yielding a convex inner approximation that retains the current feasible point and separates across cells:
\begin{subequations}\label{eq:qsub}
\begin{align}
 \min\limits_{\{\mathbf q_{k,n}\}_{n\in\mathcal N_k},\delta_k}\quad
 &\delta_k^2+\frac1{2\rho}
 \sum_{n\in\mathcal N_k}\sum_{i\in\mathcal S_n}
 |q_{ki,n}-t_{ki,n}|^2
 \label{eq:qsub_obj}\\
 \mathrm{s.t.}\quad
 &d_k-\delta_k-
 \sum_{n\in\mathcal N_k}\widetilde r_{k,n}(\mathbf q_{k,n})\le0,
 \label{eq:qsub_service}\\
 &0\le\delta_k\le\bar\delta_k,
 \label{eq:qsub_bounds}
\end{align}
\end{subequations}
where $t_{ki,n}$ denotes the entry of $\mathbf T_n$ indexed by $(k,i)$. Constraint~\eqref{eq:qsub_service} requires the sum of the slotwise lower bounds for cell $k$ to be at least $d_k-\delta_k$. 

We solve \eqref{eq:qsub} via the Karush-Kuhn-Tucker (KKT) conditions. Let $\pi_k\ge0$ denote the multiplier of \eqref{eq:qsub_service}. For fixed $\pi_k$, the stationarity condition with respect to $q_{ki,n}^{*}$ yields
\begin{equation}
 q_{ki,n}^{\star}(\pi_k)
 =\frac{t_{ki,n}+2\rho\pi_k b_{k,n}\mathbf1_{\{i=k\}}}
 {1+2\rho\pi_k a_{k,n}},
 \label{eq:qsol}
\end{equation}
and the minimization of $\delta_k^2-\pi_k\delta_k$ over \eqref{eq:qsub_bounds} yields
\begin{equation}
 \delta_k^{\star}(\pi_k)=[\pi_k/2]_0^{\bar\delta_k}.
 \label{eq:deltasol}
\end{equation}
Setting $\pi_k=0$ gives the candidate $\mathbf q_{k,n}=\mathbf t_{k,n}\triangleq[t_{ki,n}]_{i\in\mathcal S_n}^{\mathrm T}$ and $\delta_k=0$. If this candidate satisfies \eqref{eq:qsub_service}, it attains the minimum objective value of zero, so $\pi_k^{\star}=0$. Otherwise, complementary slackness requires $\Psi_k(\pi_k^{\star})=0$, where $\Psi_k(\pi_k)=d_k-\delta_k^{\star}(\pi_k)-\sum_{n\in\mathcal N_k}\widetilde r_{k,n}\bigl(\mathbf q_{k,n}^{\star}(\pi_k)\bigr)$ is the residual of \eqref{eq:qsub_service}. As $\pi_k$ increases, $\mathbf q_{k,n}^{\star}(\pi_k)$ in \eqref{eq:qsol} moves from $\mathbf t_{k,n}$ toward the maximizer of the concave bound $\widetilde r_{k,n}$ (or stays at $\mathbf t_{k,n}$ if $a_{k,n}=0$), and $\delta_k^{\star}(\pi_k)$ in \eqref{eq:deltasol} is nondecreasing; hence, $\Psi_k$ is continuous and nonincreasing. Under Slater's condition, $\Psi_k(\pi_k)<0$ for sufficiently large $\pi_k$. Thus, $\pi_k^{\star}$ can be bracketed by doubling a trial multiplier and then located by bisection, and evaluating \eqref{eq:qsol} and \eqref{eq:deltasol} at $\pi_k^{\star}$ gives the unique solution of \eqref{eq:qsub}.
Solving \eqref{eq:qsub} once for each cell completes one feasible descent update of $(\mathbf Q,\boldsymbol\delta)$ within the BCD cycle. The receive coefficients, weights, and lower bounds on $r_{k,n}$ are refreshed at the latest $\mathbf Q$ in the next cycle.

\subsubsection{Subproblem with respect to $\mathbf W$}
With $\mathbf Q$ and $\mathbf Z$ fixed, beamforming optimization reduces to a power-constrained least-squares problem with target $\mathbf Y_n=\mathbf Q_n+\rho\mathbf\Lambda_n$. After scaling the $\mathbf W$-dependent part of \eqref{eq:AL} by $2\rho$ and adding a proximal term, the slotwise subproblem can be expressed as
\begin{subequations}\label{eq:wsub}
\begin{align}
\min\limits_{\mathbf W_n}\quad
 &\|\mathbf H_n\mathbf W_n-\mathbf Y_n\|_F^2+\tau_W\|\mathbf W_n-\mathbf W_n^{(j)}\|_F^2
 \label{eq:wsub_obj}\\
 \mathrm{s.t.}\quad&\|\mathbf W_n\|_F^2\le P_{\mathrm s},
 \label{eq:wsub_power}
\end{align}
\end{subequations}
where $\tau_W>0$ ensures strong convexity and a unique minimizer~\cite{Razaviyayn2013BSUM}. Since the proximal term is nonnegative and vanishes at the current feasible point $\mathbf W_n^{(j)}$, the update preserves descent of the AL objective.

We solve \eqref{eq:wsub} via its KKT conditions. With multiplier $\zeta_n\ge0$ for \eqref{eq:wsub_power}, $\mathbf A_n=\mathbf H_n^{\mathrm H}\mathbf H_n+\tau_W\mathbf I_M\succ\mathbf0$, and $\mathbf V_n=\mathbf H_n^{\mathrm H}\mathbf Y_n+\tau_W\mathbf W_n^{(j)}$, the stationarity condition gives
\begin{equation}
 \mathbf W_n^{\star}(\zeta_n)=(\mathbf A_n+\zeta_n\mathbf I_M)^{-1}\mathbf V_n.
 \label{eq:wsol}
\end{equation}
Let $P_n(\zeta_n)=\|\mathbf W_n^{\star}(\zeta_n)\|_F^2$. If $P_n(0)\le P_{\mathrm s}$, then $\zeta_n^{\star}=0$. Otherwise, complementary slackness requires $P_n(\zeta_n)=P_{\mathrm s}$. Since $P_n$ is continuous and strictly decreasing, this equation has a unique positive solution $\zeta_n^\star$, which we find by bisection on $[0,\|\mathbf V_n\|_F/\sqrt{P_{\mathrm s}}]$ using a single eigendecomposition per slot.

\subsubsection{Subproblem with respect to $\mathbf Z$}
For fixed $\mathbf Q$ and $\mathbf W$, multiplying the position-dependent part of the AL by $\rho>0$ gives
\begin{subequations}\label{eq:zoriginal}
\begin{align}
 \min\limits_{\mathbf Z}\quad
 &F_Z(\mathbf Z)=\frac12\sum_n
 \|\mathbf H_n(\mathbf Z_n)\mathbf W_n-\mathbf Y_n\|_F^2
 \label{eq:zoriginal_obj}\\
 \mathrm{s.t.}\quad
 &\eqref{prob:P0_region}\text{-}\eqref{prob:P0_speed}.
 \label{eq:zoriginal_constraints}
\end{align}
\end{subequations}

To address the nonconvex objective and minimum-spacing constraints of problem \eqref{eq:zoriginal}, we use SCA to construct a convex subproblem at each iteration. The antenna positions are optimized jointly across the entire service period because the movement-speed constraints couple adjacent slots.

The gradient of $F_Z$ with respect to $\mathbf z_{m,n}$ can be expressed as
\begin{equation}
 \nabla_{\mathbf z_{m,n}}F_Z
 =-\sum_{k,i\in\mathcal S_n}\Im\left\{
 E_{ki,n}^*[\mathbf H_n]_{k,m}w_{mi,n}
 \right\}\mathbf a_{k,n},
 \label{eq:zgrad}
\end{equation}
where $E_{ki,n}=s_{ki,n}-y_{ki,n}$ is the effective-signal error, $y_{ki,n}=[\mathbf Y_n]_{k,i}$, $w_{mi,n}$ denotes the $m$-th entry of the beamforming vector for illuminated cell $i$, the normalized channel entry is $[\mathbf H_n(\mathbf Z_n)]_{k,m}=\vartheta_{k,n}e^{j\mathbf a_{k,n}^{\mathrm T}\mathbf z_{m,n}}$, $\vartheta_{k,n}=\frac{\sqrt{\beta_{k,n}}}{\sigma_k}e^{-j2\pi\|\mathbf d_{k,n}\|_2/\lambda}$, and the effective signal is $s_{ki,n}=\sum_m\vartheta_{k,n}w_{mi,n}e^{j\mathbf a_{k,n}^{\mathrm T}\mathbf z_{m,n}}$.

At each BCD iteration, we initialize an inner SCA loop at the current geometrically feasible positions, with $s$ denoting its iteration index. At the expansion point $\mathbf Z^{(s)}$, we define $\mathbf g_{m,n}^{(s)}=\nabla_{\mathbf z_{m,n}}F_Z(\mathbf Z^{(s)})$ and $\mathbf d_{mm',n}^{(s)}=\mathbf z_{m,n}^{(s)}-\mathbf z_{m',n}^{(s)}$. The following lemma provides a quadratic upper bound for $F_Z$.

\begin{lemma}\label{lem:curvature}
For fixed $\mathbf W$ and $\mathbf Y$, let $\mathbf c_{ki,n}=\vartheta_{k,n}[w_{1i,n},\ldots,w_{Mi,n}]^{\mathrm T}$ and
\begin{equation}
\begin{aligned}
 L_n={}&\sum_{k,i\in\mathcal S_n}\|\mathbf a_{k,n}\|_2^2
 \|\mathbf c_{ki,n}\|_2^2\\
 &+\sum_{k,i\in\mathcal S_n}\|\mathbf a_{k,n}\|_2^2
 (\|\mathbf c_{ki,n}\|_1+|y_{ki,n}|)
 \|\mathbf c_{ki,n}\|_\infty.
\end{aligned}
\label{eq:Lz}
\end{equation}
Then, $\|\nabla_{\operatorname{vec}(\mathbf Z_n)}^2F_Z\|_2\le L_n$ holds for all $\mathbf Z$ and $n\in\mathcal N$. Consequently, for any $\mathbf Z$,
\begin{equation}
\begin{aligned}
 F_Z(\mathbf Z)\le{}&F_Z(\mathbf Z^{(s)})+\sum_{n,m}(\mathbf g_{m,n}^{(s)})^{\mathrm T}
 (\mathbf z_{m,n}-\mathbf z_{m,n}^{(s)})\\
 &+\frac12\sum_nL_n\|\mathbf Z_n-\mathbf Z_n^{(s)}\|_F^2.
\end{aligned}
\label{eq:zmaj}
\end{equation}
\end{lemma}
\begin{IEEEproof}
See Appendix~\ref{app:curvature}.
\end{IEEEproof}

The first-order lower bound of the squared spacing gives the convex inner approximation
\begin{equation}
\begin{aligned}
 2(\mathbf d_{mm',n}^{(s)})^{\mathrm T}
 (\mathbf z_{m,n}-\mathbf z_{m',n})-\|\mathbf d_{mm',n}^{(s)}\|_2^2\ge d_{\min}^2.
\end{aligned}
\label{eq:spacinglin}
\end{equation}

Using \eqref{eq:zmaj} and \eqref{eq:spacinglin}, the SCA subproblem can be expressed as
\begin{subequations}\label{eq:zsub}
\begin{align}
 \min\limits_{\mathbf Z}\quad
 &\sum_{n,m}(\mathbf g_{m,n}^{(s)})^{\mathrm T}
 (\mathbf z_{m,n}-\mathbf z_{m,n}^{(s)})\notag\\
 &+\frac12\sum_n L_n
 \|\mathbf Z_n-\mathbf Z_n^{(s)}\|_F^2
 \label{eq:zsub_obj}\\
 \mathrm{s.t.}\quad
 &\mathbf z_{m,n}\in\mathcal A,\quad\forall m,n,
 \label{eq:zsub_region}\\
 &\eqref{eq:spacinglin},\quad\forall m<m',n,
 \label{eq:zsub_spacing}\\
 &\|\mathbf z_{m,n}-\mathbf z_{m,n-1}\|_2\le v_{\max}\Delta t,\notag\\
 &\hspace{20mm}\forall m,\ n=2,\ldots,N.
 \label{eq:zsub_speed}
\end{align}
\end{subequations}
For the rectangular MA region used in this work, \eqref{eq:zsub} is a convex quadratically constrained quadratic program (QCQP). It can be reformulated as a second-order cone program (SOCP) and solved by CVX~\cite{GrantBoyd2008CVX}. The resulting solution gives $\mathbf Z^{(s+1)}$, at which the gradients and spacing approximations are recomputed for the next SCA iteration. With the other blocks fixed, these SCA iterations terminate when
\begin{equation}
 \left|F_Z(\mathbf Z^{(s+1)})-F_Z(\mathbf Z^{(s)})\right|
 \le\epsilon_Z,
 \label{eq:z_stopping}
\end{equation}
where $\epsilon_Z>0$ is the prescribed SCA stopping tolerance for the subproblem with respect to $\mathbf Z$.

\subsection{Overall Algorithm}

\subsubsection{Initialization and outer updates}
For the BH pattern returned by Algorithm~\ref{alg:DAAS}, PDD is initialized with the feasible reference transmission in \eqref{eq:refbeam}-\eqref{eq:refrate}, i.e., $\mathbf W^{(0)}=\mathbf W^{\mathrm r}$, $\mathbf Z^{(0)}=\mathbf Z^{\mathrm r}$, $\mathbf Q_n^{(0)}=\mathbf H_n(\mathbf Z_n^{\mathrm r})\mathbf W_n^{\mathrm r}$, and $\delta_k^{(0)}=[d_k-\sum_{n\in\mathcal N_k}r_{k,n}^{\mathrm r}(\mathcal S_n)]_+$. At outer iteration $t$, BCD starts from $\mathcal V^{t,0}=\mathcal V^t$ with $\rho_t$ and $\{\mathbf\Lambda_n^t\}$ fixed, and terminates after a complete cycle $j\ge1$ when
\begin{equation}
 \left|\mathcal L_{\rho_t}(\mathcal V^{t,j})
 -\mathcal L_{\rho_t}(\mathcal V^{t,j-1})\right|
 \le\epsilon_{\mathrm{in}},
 \label{eq:bcd_stopping}
\end{equation}
where $\epsilon_{\mathrm{in}}>0$ is the BCD tolerance. Then, set $\mathcal V^{t+1}=\mathcal V^{t,j}$ and compute the consistency residual $\mathcal E_t=\max_n\|\mathbf R_n^{t+1}\|_{\max}$, with $\mathcal E_{-1}=+\infty$. Following the residual reduction rule in~\cite{Xue2026PASS}, if $\mathcal E_t\le0.8\mathcal E_{t-1}$, the multipliers are updated as $\mathbf\Lambda_n^{t+1}=\mathbf\Lambda_n^t+\mathbf R_n^{t+1}/\rho_t$ with $\rho_{t+1}=\rho_t$; otherwise, $\mathbf\Lambda_n^{t+1}=\mathbf\Lambda_n^t$ and $\rho_{t+1}=0.8\rho_t$. The outer loop terminates when $\mathcal E_t\le\epsilon_{\rm con}$. Finally, $C_k$ is evaluated using the final beams and channels, and the beam powers are recovered via Appendix~\ref{app:reach} to attain $C_k^{\mathrm{tar}}=\min\{C_k,D_k\}$. The two-stage procedure is summarized in Algorithm~\ref{alg:pdd}.

\begin{algorithm}[t]
\DontPrintSemicolon
\SetAlgoNoLine
\SetKwInOut{KwIn}{Input}
\SetKwInOut{KwOut}{Output}
\SetNlSty{textnormal}{}{:}
\caption{Two-Stage MA-BH Optimization}
\label{alg:pdd}
\KwIn{$\mathcal K,\allowbreak\mathcal N,\allowbreak M,\allowbreak\mathcal A,\allowbreak\mathbf Z^{\mathrm r};\allowbreak P_{\mathrm s},\allowbreak B_{\max},\allowbreak d_{\min},\allowbreak v_{\max},\allowbreak\Delta t,\allowbreak\kappa,\allowbreak\eta_{\mathrm{sp}};\allowbreak\{\mathbf a_{k,n},\beta_{k,n},\vartheta_{k,n}\}_{k,n};\allowbreak\{d_k,c_k^{\min},\sigma_k^2\}_k;\allowbreak\rho_0,\allowbreak\tau_W,\allowbreak\epsilon_Z,\allowbreak\epsilon_{\mathrm{in}},\allowbreak\epsilon_{\rm con},\allowbreak\epsilon_{\mathrm{match}}$.}
Obtain $\mathbf X$ from Algorithm~\ref{alg:DAAS} and fix it for PDD\;
Initialize $\mathcal V^0$ as above; set $\mathbf\Lambda_n^0=\mathbf0$, $\mathcal E_{-1}=+\infty$, and $t\gets0$\;
\Repeat{$\mathcal E_{t-1}\le\epsilon_{\rm con}$}{
 Fix $\rho_t,\{\mathbf\Lambda_n^t\}_n$; set $\mathcal V^{t,0}=\mathcal V^t$ and $j\gets0$\;
 \Repeat{\eqref{eq:bcd_stopping} holds}{
  Refresh $u_{k,n},v_{k,n}$ and update $(\mathbf Q,\boldsymbol\delta)$ via \eqref{eq:qsub}\;
  Update $\mathbf W$ by \eqref{eq:wsol}\;
  Update $\mathbf Z$ by SCA solves of \eqref{eq:zsub} until \eqref{eq:z_stopping} holds\;
  Set $j\gets j+1$ and store the updated variables as $\mathcal V^{t,j}$\;
 }
 Set $\mathcal V^{t+1}=\mathcal V^{t,j}$, compute $\mathcal E_t$, and update $\{\mathbf\Lambda_n^{t+1}\}$ and $\rho_{t+1}$ by the residual reduction rule\;
 $t\gets t+1$\;
}
Evaluate $C_k$ using the final $(\mathbf W,\mathbf Z)$ and recover the final beam powers via Appendix~\ref{app:reach}\;
\KwOut{$\mathbf X,\mathbf W,\mathbf Z$.}
\end{algorithm}

\subsubsection{Convergence}
Under the assumptions on $\underline r_k$ and the initial quotas, the slotwise construction in Algorithm~\ref{alg:DAAS} yields a feasible reference transmission within $NB_{\max}$ cell selections. Its matching operations preserve this feasibility and $C_\omega(\mathbf X)\le C_0$, accepting at most $\lfloor\Phi(\mathbf X^{(0)})/\epsilon_{\mathrm{match}}\rfloor$ moves. At termination, no admissible replacement or swap decreases $\Phi$ by at least $\epsilon_{\mathrm{match}}$.
For fixed $\rho_t$ and $\{\mathbf\Lambda_n^t\}$, the WMMSE lower bounds on $r_{k,n}$ are tight at the current feasible point. Solving the $(\mathbf Q,\boldsymbol\delta)$ inner approximation therefore preserves feasibility of the AL subproblem and does not increase its objective. The proximal $\mathbf W$ and SCA $\mathbf Z$ updates also preserve descent. Since $\mathcal L_{\rho_t}\ge0$, the inner AL values form a convergent nonincreasing sequence.
If the consistency residuals and the first-order errors of the inner AL solutions vanish, the primal iterates and effective multipliers $\{\mathbf\Lambda_n^t+\mathbf R_n^{t+1}/\rho_t\}_n$ remain bounded, then any accumulation point of the PDD iterates at which Robinson's condition holds satisfies the KKT conditions of $\mathcal P_\delta$ with $\mathbf X$ fixed~\cite{Shi2020PDD1}. In the implementation, \eqref{eq:bcd_stopping} and $\mathcal E_t\le\epsilon_{\rm con}$ serve as the stopping criteria for the inner BCD and outer PDD loops, respectively.

\subsubsection{Complexity analysis}
For a rectangular aperture, computing the conflict coefficients and reference correlations costs $\mathcal O(NMK^2)$, and sorting the interferers costs $\mathcal O(NK^2\log K)$. The slotwise construction requires $\mathcal O(NKB_{\max})$ selection operations. With cached cumulative service and per-slot conflict sums, each of the $Q_{\mathrm{sw}}$ candidate moves in the matching search costs $\mathcal O(K+B_{\max}^2)$, where a full replacement scan examines at most $L(KN-L)$ candidates with $L=\sum_{k,n}x_{k,n}^{(0)}$, and a full swap scan at most $\binom N2 B_{\max}^2$. Hence, DAAS avoids enumerating the $\bigl(\sum_{b=0}^{B_{\max}}\binom Kb\bigr)^N$ possible patterns over the service period. With $I_{\rm out}$ PDD iterations, $I_{\rm BCD}$ BCD cycles per PDD iteration, $I_Z$ SCA iterations per $\mathbf Z$ update, and scalar-search counts $J_Q$ and $J_W$, the overall complexity is
\begin{equation}
\begin{aligned}
 \mathcal O\Bigl(&NK^2(M+\log K)+Q_{\mathrm{sw}}(K+B_{\max}^2)+NB_{\max}^3\\
 &+I_{\rm out}I_{\rm BCD}\bigl[NB_{\max}^2J_Q\\
 &+NM\bigl(M^2+B_{\max}(M+B_{\max}+J_W)\bigr)\\
 &+I_Z\bigl(NM(B_{\max}^2+M)+\mathcal T_{\rm socp}\bigr)\bigr]\Bigr),
\end{aligned}
\label{eq:complexity}
\end{equation}
where $\mathcal T_{\rm socp}$ is the cost of one SOCP solve for $\mathbf Z$ over all slots, $NB_{\max}^3$ accounts for the final power recovery, and the $\mathcal O(NKB_{\max})$ selection cost is dominated by $NK^2M$ since $B_{\max}\le K$. Fixing the pattern reduces the signal auxiliaries from $NK^2$ to $\sum_n b_n^2\le NB_{\max}^2$ and eliminates the binary BH indicators from the continuous stage.

\section{Simulation Results}
\label{sec:simulations}

\subsection{Simulation Setup}
We consider a satellite in a typical Walker Delta LEO constellation with $R_{\mathrm e}=6{,}371$~km, $\mu=3.9860\times10^{14}$~m$^3$/s$^2$, and $c=3\times10^8$~m/s, serving a fixed circular region centered directly beneath it at the midpoint of the service period. Cell centers are sampled uniformly over the region's surface area, and $\mathbf Z^{\mathrm r}$ is set to a $4\times4$ uniform planar array (UPA) centered in $\mathcal A$ with $\lambda/2$ spacing along the SCCS $x$- and $y$-axes, where $\lambda=c/f_c$. Unless otherwise stated, the default system parameters~\cite{Zhu2025DynamicMA,3GPP2026NTNRF} and the remaining algorithm parameters are listed in Table~\ref{tab:sim_parameters}.

\begin{table}[t]
 \centering
 \caption{Simulation Parameters}
 \label{tab:sim_parameters}
 \footnotesize
 \renewcommand{\arraystretch}{1.12}
 \setlength{\tabcolsep}{3pt}
 \begin{tabular}{|>{\raggedright\arraybackslash}m{0.52\columnwidth}|>{\raggedright\arraybackslash}m{0.42\columnwidth}|}
  \hline
  \textbf{Parameter} & \textbf{Value} \\
  \hline
  Orbital altitude, $H_{\mathrm s}$ & $1200$~km \\
  \hline
  Orbital inclination, $\beta$ & $65^{\circ}$ \\
  \hline
  Initial orbital phase, $\alpha_0$ & $-7.5^{\circ}$ \\
  \hline
  Service period, $T_H$ & $273.17$~s \\
  \hline
  Number of slots, $N$ & $50$ \\
  \hline
  Number of cells, $K$ & $8$ \\
  \hline
  Region angular radius & $6^{\circ}$ \\
  \hline
  Carrier frequency, $f_c$ & $20$~GHz \\
  \hline
  Bandwidth, $W$ & $200$~MHz \\
  \hline
  Number of MA elements, $M$ & $16$ \\
  \hline
  Maximum illuminated cells, $B_{\max}$ & $4$ \\
  \hline
  Total transmit-power budget, $P_{\mathrm s}$ & $100$~W \\
  \hline
  Satellite antenna gain, $G_{\mathrm s}$ & $0$~dBi \\
  \hline
  Receive-antenna gain, $G_k$ & $42$~dBi \\
  \hline
  Noise power, $\sigma_k^2$ & $-120$~dBW \\
  \hline
  MA movement region, $\mathcal A$ & $[-1.5\lambda,1.5\lambda]^2$ \\
  \hline
  Minimum element spacing, $d_{\min}$ & $\lambda/2$ \\
  \hline
  Maximum MA speed, $v_{\max}$ & $0.0074$~m/s \\
  \hline
  Cumulative traffic demand, $D_k$ & $\mathcal U[20,80]$~Gbit \\
  \hline
  Minimum delivered data, $C_k^{\min}$ & $5$~Gbit \\
  \hline
  Algorithm parameters & $\eta_{\mathrm{sp}}=1$, $\rho_0=10^{-3}$, $\tau_W=10^{-3}$, $\epsilon_Z=10^{-5}$, $\epsilon_{\mathrm{in}}=10^{-3}$, $\epsilon_{\rm con}=10^{-4}$ \\
  \hline
 \end{tabular}
\end{table}

\subsection{Benchmark Schemes}
To evaluate the performance of the proposed \textbf{MA-DAAS-PDD} scheme, we consider the following benchmarks from the perspectives of antenna array type and BH pattern design.
\begin{itemize}
 \item \textbf{UPA-DAAS-PDD:} The MA array is replaced by the fixed reference UPA, and PDD optimizes only the beamforming vectors under the BH pattern returned by Algorithm~\ref{alg:DAAS}.
 \item \textbf{LCMA-DAAS-PDD:} To reduce the repositioning overhead, this low-complexity MA (LCMA) scheme optimizes a common set of MA positions $\mathbf z_{m,n}=\mathbf z_m$ using the first $20$ slots, keeps it fixed throughout the service period, and optimizes the slot-dependent beamforming vectors under the same BH pattern.
 \item \textbf{MA-Relaxed-PDD:} The binary constraint is relaxed to $0\le x_{k,n}\le1$~\cite{Wang2024JointBIP}, and $\mathbf X$, $\mathbf W$, and $\mathbf Z$ are iteratively updated by BCD within a unified PDD framework, with a penalty on $\sum_{k,n}x_{k,n}(1-x_{k,n})$ driving $x_{k,n}$ toward $0$ or $1$. A binary pattern is finally recovered subject to the quotas $\ell_{k,1}$ and per-slot illumination limits.
 \item \textbf{MA-RR-PDD:} Cells are illuminated in a fixed RR order~\cite{AlegreGodoy2012OfferedCapacity} with at most $B_{\max}$ cells per slot, and the PDD stage of Algorithm~\ref{alg:pdd} optimizes the beamforming vectors and MA positions.
\end{itemize}

\subsection{Convergence of the Proposed Algorithm}
\begin{figure*}[!t]
    \centering
    % Common image height; the minipage widths follow the image aspect ratios
    % (fig2: 252x241 pt, fig4/fig5: 4:3), so all three figures have equal heights.
    % Set outside the minipages because \textwidth is redefined inside them.
    \setlength{\figrowheight}{0.26\textwidth}
    \begin{minipage}[t]{0.272\textwidth}
        \vspace{0pt}
        \centering
        \includegraphics[height=\figrowheight]{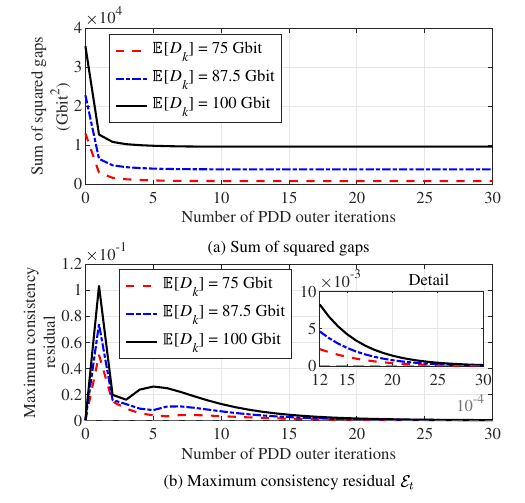}
        \caption{Convergence behaviors of the proposed DAAS-PDD algorithm.}
        \label{fig:convergence}
    \end{minipage}\hfill
    \begin{minipage}[t]{0.347\textwidth}
        \vspace{0pt}
        \centering
        \includegraphics[height=\figrowheight]{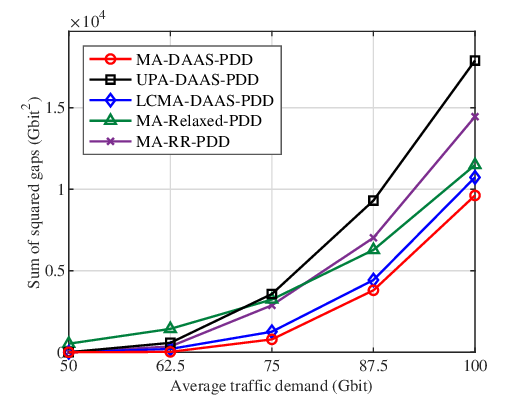}
        \caption{Sum of squared gaps versus the average traffic demand for the proposed and benchmark schemes.}
        \label{demand_sweep12}
    \end{minipage}\hfill
    \begin{minipage}[t]{0.347\textwidth}
        \vspace{0pt}
        \centering
        \includegraphics[height=\figrowheight]{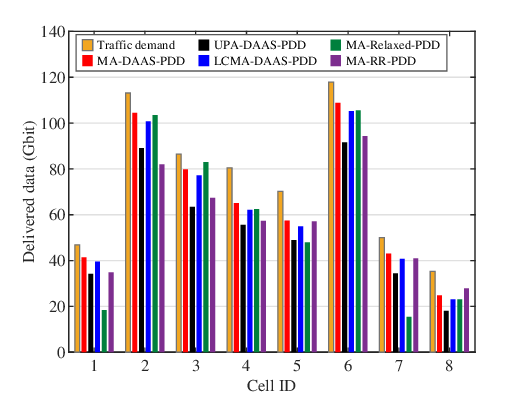}
        \caption{Per-cell delivered data and traffic demand with $\mathbb E[D_k] = 75$~Gbit.}
        \label{cell_service}
    \end{minipage}
\end{figure*}

Figure~\ref{fig:convergence}(a) and (b) show the sum of squared gaps and the maximum consistency residual $\mathcal E_t$ of the DAAS-PDD algorithm for average traffic demands of $75$, $87.5$, and $100$~Gbit. The sum of squared gaps decreases rapidly in the first few outer iterations and nearly stabilizes after ten iterations, reaching $789.46$, $3800.77$, and $9619.03$~Gbit$^2$, respectively, since higher demands leave larger gaps under the same resources. The residual starts at zero because the reference initialization satisfies \eqref{eq:qcons}, increases as intermediate PDD updates break this equality, and falls below $\epsilon_{\mathrm{con}}=10^{-4}$ after $25$, $27$, and $30$ iterations, respectively. Higher demands require more outer iterations, since reducing larger shortfalls requires larger adjustments of the auxiliary signal coefficients and thus produces larger transient residuals. Hence, most of the demand matching improvement occurs early, whereas the remaining iterations mainly enforce consistency between the auxiliary variables and the physical transmission.

\subsection{Performance Comparison With Benchmark Schemes}

Figure~\ref{demand_sweep12} shows the sum of squared gaps versus the average traffic demand from $50$ to $100$~Gbit. The gaps grow with the traffic demand under fixed transmission resources, while the proposed \textbf{MA-DAAS-PDD} achieves the smallest gaps throughout the tested range.
Compared with \textbf{UPA-DAAS-PDD} under the same BH pattern, \textbf{MA-DAAS-PDD} additionally adjusts the antenna positions to reshape the inter-cell channel correlation, thereby separating simultaneously illuminated cells that the fixed UPA can hardly separate by beamforming. 
\textbf{LCMA-DAAS-PDD} also benefits from optimized positions, but its common array cannot follow the time-varying illumination requirements of BH, i.e., the illuminated-cell sets and viewing geometry that change across slots, whereas \textbf{MA-DAAS-PDD} adjusts its positions slot by slot. Nevertheless, LCMA reduces the position-optimization cost and movement overhead, offering a performance-complexity tradeoff.
The advantage over \textbf{MA-RR-PDD} comes from the BH pattern generation: RR illumination ignores both the residual demands and the spatial separability of simultaneously illuminated cells, whereas DAAS accounts for both under the minimum-service quotas and directs illumination opportunities toward cells with larger shortfalls. For \textbf{MA-Relaxed-PDD}, recovering binary decisions from the relaxed solution can change its interference and power allocation, so a favorable fractional design need not yield a favorable binary pattern. In contrast, DAAS directly constructs a binary pattern by accounting for the cell demands and the correlation reduction achievable by MA movement, making the simultaneous illumination of neighboring cells less prone to inter-beam interference.

Figure~\ref{cell_service} compares the delivered data and traffic demand of each cell with $\mathbb E[D_k] = 75$~Gbit. We can observe that \textbf{MA-DAAS-PDD} outperforms \textbf{UPA-DAAS-PDD} across all cells, with the largest gains for the highest-demand cells $2$, $3$, and $6$; e.g., the delivered data of cells $2$ and $6$ rise from $89.11$ and $91.58$~Gbit to $104.47$ and $108.84$~Gbit, respectively. Although \textbf{MA-RR-PDD} and \textbf{MA-Relaxed-PDD} deliver more data to cells $8$ and $3$, respectively, both exhibit larger sums of squared gaps. This indicates that higher data delivery to individual cells does not necessarily translate into better system-wide demand matching. By accounting for residual demands and the conflicts that MA movement cannot remove, DAAS coordinates illumination across cells, reducing large service shortfalls and achieving more accurate demand matching.

\subsection{Impact of System Parameters}
\begin{figure*}[!t]
    \centering
    \begin{minipage}[t]{0.325\textwidth}
        \vspace{0pt}
        \centering
        \includegraphics[width=\linewidth]{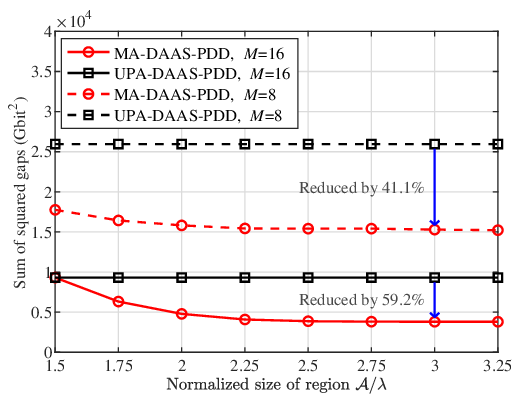}
        \caption{Sum of squared gaps versus the normalized side length of the MA movement region.}
        \label{aperture_sweep}
    \end{minipage}\hfill
    \begin{minipage}[t]{0.325\textwidth}
        \vspace{0pt}
        \centering
        \includegraphics[width=\linewidth]{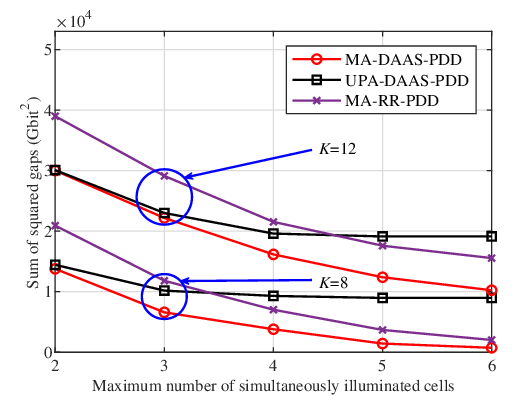}
        \caption{Sum of squared gaps versus the maximum number of simultaneously illuminated cells.}
        \label{beam_limit_sweep}
    \end{minipage}\hfill
    \begin{minipage}[t]{0.325\textwidth}
        \vspace{0pt}
        \centering
        \includegraphics[width=\linewidth]{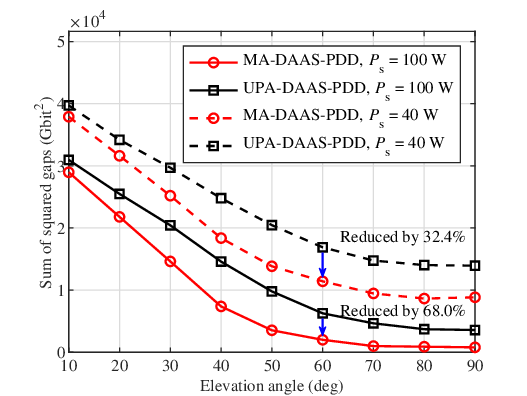}
        \caption{Sum of squared gaps versus the elevation angle.}
        \label{elevation_sweep}
    \end{minipage}
\end{figure*}

Figure~\ref{aperture_sweep} illustrates the sum of squared gaps versus the normalized side length of the square MA movement region for $\mathbb E[D_k] =87.5$~Gbit. The UPA curves remain constant due to their fixed antenna positions, whereas the MA curves decrease initially before gradually stabilizing. Enlarging the movement region provides greater flexibility in adjusting the relative phases across cell directions, thereby improving the separation of simultaneous transmissions. At a side length of $3\lambda$, MA reduces the sum of squared gaps by $41.1\%$ and $59.2\%$ compared with UPA for $M=8$ and $M=16$, respectively. However, the gains become marginal beyond approximately $2.5\lambda$. These results demonstrate that the flexibility of MA position adjustment, together with the antenna count, enhances system-wide demand matching, although further enlarging the movement region yields diminishing returns.

Figure~\ref{beam_limit_sweep} shows the sum of squared gaps versus $B_{\max}$ for $\mathbb E[D_k]=87.5$~Gbit. It can be observed that the sums are larger for $K=12$ than for $K=8$ because more cells compete for the same resources. As $B_{\max}$ increases, the sums of the MA-based schemes keep decreasing, whereas \textbf{UPA-DAAS-PDD} saturates beyond $B_{\max}=4$, because inter-beam interference prevents a fixed array from exploiting the additional illumination allowance. Optimizing the MA positions alleviates this interference and allows more neighboring cells to be illuminated simultaneously, so the gap between \textbf{MA-DAAS-PDD} and \textbf{UPA-DAAS-PDD}, which is small at $B_{\max}=2$, widens as $B_{\max}$ increases. \textbf{MA-RR-PDD} is inferior to \textbf{UPA-DAAS-PDD} at small $B_{\max}$ and outperforms it only at larger $B_{\max}$, while \textbf{MA-DAAS-PDD} achieves the lowest sum for all $B_{\max}$. This shows that both MA position optimization and demand-aware pattern design are needed to benefit from a larger $B_{\max}$. It is worth noting that a larger $B_{\max}$ imposes a heavier signal-processing burden on the payload.

Figure~\ref{elevation_sweep} shows the sum of squared gaps versus the elevation angle, measured at the region center at the midpoint of the service period, for $\mathbb E[D_k]=75$~Gbit. Both schemes improve with increasing elevation and flatten at high elevations, since a shorter propagation distance reduces the free-space path loss. \textbf{MA-DAAS-PDD} outperforms \textbf{UPA-DAAS-PDD} at every tested elevation and power budget, with a modest advantage at low elevations, where the large propagation loss limits the benefit of interference control, and a more pronounced one at moderate and high elevations. For example, at $60^{\circ}$, \textbf{MA-DAAS-PDD} reduces the sum of squared gaps by $68.0\%$ and $32.4\%$ for $P_{\mathrm s}=100$ and $40$~W, respectively. Increasing $P_{\mathrm s}$ reduces the gaps for both schemes, while the persistent MA advantage shows that position optimization provides gains beyond those of a higher transmit power.

\subsection{MA Positions and Beam Patterns}
\begin{figure*}[t!]
    \centering
    \captionsetup[subfloat]{font=footnotesize,labelfont=rm,textfont=rm,skip=2pt}
    \subfloat[Antenna positions, $n=28$\label{array_geometry28}]{\includegraphics[width=0.247\textwidth]{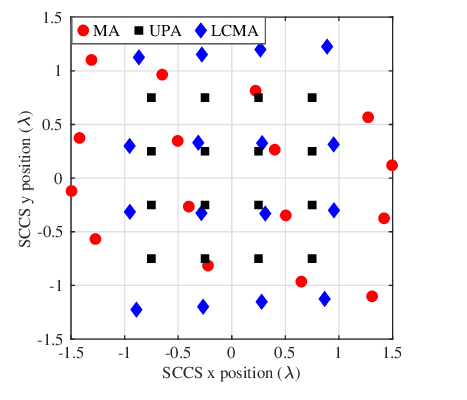}}\hfill
    \subfloat[Dynamic MA, $n=28$\label{beam_gain_slot_28_dynamic_ma}]{\includegraphics[width=0.247\textwidth]{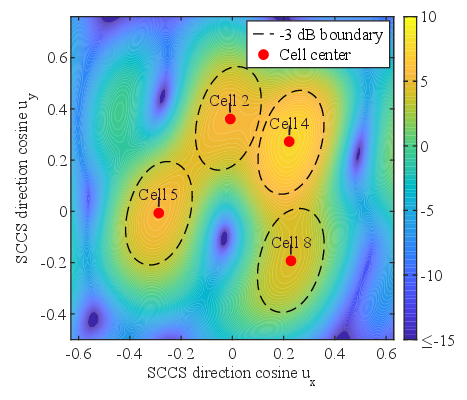}}\hfill
    \subfloat[LCMA, $n=28$\label{beam_gain_slot_28_lcma}]{\includegraphics[width=0.247\textwidth]{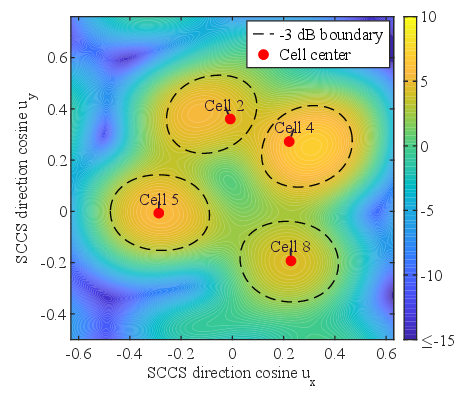}}\hfill
    \subfloat[UPA, $n=28$\label{beam_gain_slot_28_upa}]{\includegraphics[width=0.247\textwidth]{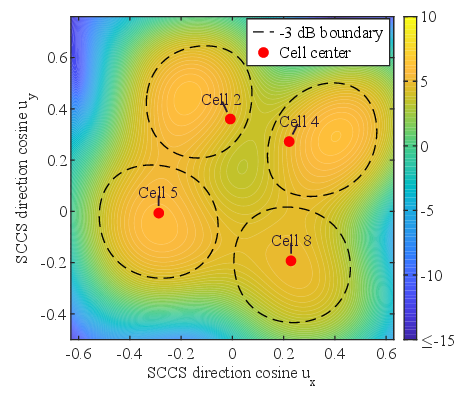}}
    \par\vspace{1pt}
    \subfloat[Antenna positions, $n=40$\label{array_geometry40}]{\includegraphics[width=0.247\textwidth]{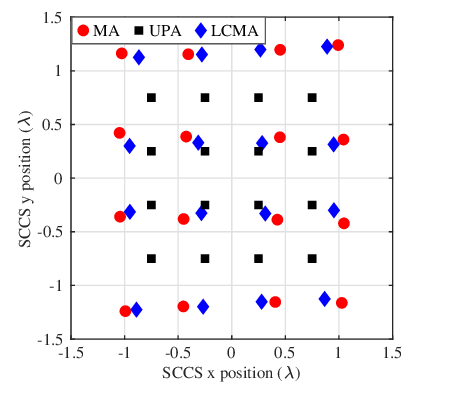}}\hfill
    \subfloat[Dynamic MA, $n=40$\label{beam_gain_slot_40_dynamic_ma}]{\includegraphics[width=0.247\textwidth]{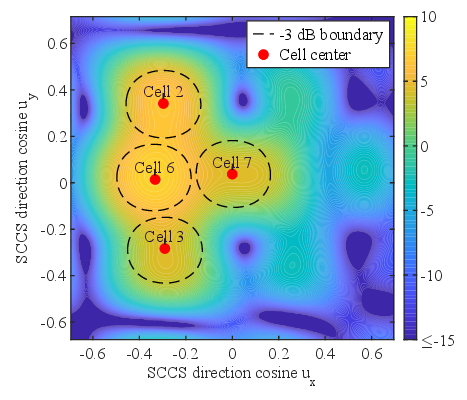}}\hfill
    \subfloat[LCMA, $n=40$\label{beam_gain_slot_40_lcma}]{\includegraphics[width=0.247\textwidth]{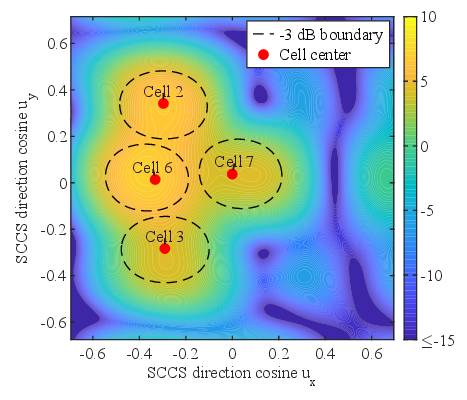}}\hfill
    \subfloat[UPA, $n=40$\label{beam_gain_slot_40_upa}]{\includegraphics[width=0.247\textwidth]{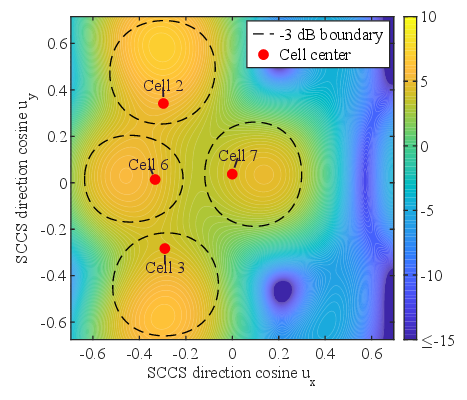}}
    \caption{Antenna positions and beam gain patterns in slots $n=28$ (top row) and $n=40$ (bottom row). The color scale shows the normalized transmit gain in dB.}
    \label{fig:array_beam_patterns}
\end{figure*}

Figure~\ref{fig:array_beam_patterns} shows the antenna positions and normalized beam gain patterns in slots $n=28$ and $n=40$ for an average traffic demand of $87.5$~Gbit, where the illuminated-cell sets are $\{2,4,5,8\}$ and $\{2,3,6,7\}$, respectively. Each dashed contour marks the $-3$~dB boundary of a beam, so a smaller contour indicates a narrower beam, and a peak closer to the red dot indicates more accurate pointing. As shown in Figs.~\ref{fig:array_beam_patterns}(a) and (e), UPA and LCMA keep the same layouts in both slots, whereas dynamic MA re-optimizes its positions for each illuminated-cell set. UPA yields the widest beams with peaks deviating from the cell centers, and LCMA yields narrower beams that remain wider than those of dynamic MA. In contrast, dynamic MA concentrates beam energy on the illuminated cells, with peaks near the cell centers and less leakage toward non-target directions, particularly in slot $n=28$, where the illuminated cells are more closely spaced. By adjusting the antenna positions according to the cells illuminated in each slot and the satellite viewing geometry, the proposed design meets the time-varying illumination requirements of BH and better exploits spatial DoFs to suppress inter-beam interference.

\section{Conclusion}
This paper proposed an MA-enabled demand-aware BH design for LEO satellite downlink communications, where the BH pattern, MA positions, and beamforming vectors were jointly optimized to minimize the sum of squared gaps between the cumulative delivered data volumes and the traffic demands of cells. A two-stage DAAS-PDD algorithm was developed, in which DAAS generated the binary BH pattern, and PDD then optimized the beamforming vectors and MA positions for the given pattern. 
Simulation results indicated that the proposed \textbf{MA-DAAS-PDD} scheme could dynamically adjust the MA positions according to the illumination requirements of each slot, thereby achieving smaller supply-demand gaps than \textbf{UPA-DAAS-PDD}. The adjusted MA positions yielded narrower beams with main lobes closer to the target cell centers and less interference leakage in non-target directions. Moreover, the proposed DAAS strategy effectively leveraged the advantages of MA to achieve more accurate demand matching than the benchmark schemes. Future work will extend the proposed design to BH scenarios under random traffic demands and cooperative multi-satellite systems.

\appendices
\section{Proof of Proposition \ref{prop:reach}}
\label{app:reach}
\renewcommand{\theequation}{A.\arabic{equation}}
\setcounter{equation}{0}
We first show that any componentwise smaller data volume target can be attained by reducing beam powers only, and then use this result to establish the equivalence. Fix a feasible design $(\mathbf X,\mathbf W,\mathbf Z)$ delivering $C_k$ bits to cell $k$ and targets $0\le C_k^{\mathrm{tar}}\le C_k$. Let $\xi_k=C_k^{\mathrm{tar}}/C_k$ for $C_k>0$ and $\xi_k=0$ otherwise. The SINR targets $\bar\gamma_{k,n}=(1+\gamma_{k,n})^{\xi_k}-1\in[0,\gamma_{k,n}]$ then satisfy $\kappa\sum_{n\in\mathcal N}\log_2(1+\bar\gamma_{k,n})=C_k^{\mathrm{tar}}$.

In slot $n$, we turn off the beams of the cells outside $\mathcal S_n^+\triangleq\{k:\bar\gamma_{k,n}>0\}$ and write $\mathbf w_{k,n}=\sqrt{p_{k,n}}\mathbf u_{k,n}$ with $\|\mathbf u_{k,n}\|_2=1$ for $k\in\mathcal S_n^+$. With the directions fixed, let $g_{k\ell,n}=|\mathbf h_{k,n}^{\mathrm T}(\mathbf Z_n)\mathbf u_{\ell,n}|^2$, where $g_{kk,n}>0$ since the original SINR is positive. The powers $\mathbf p_n'$ attaining all target SINRs satisfy
\begin{equation}
\mathbf p_n'=\mathbf F_n\mathbf p_n'+\boldsymbol\nu_n,
\label{eq:app_linear}
\end{equation}
where $[\mathbf F_n]_{k\ell}=\bar\gamma_{k,n}g_{k\ell,n}/g_{kk,n}$ for $\ell\ne k$, $[\mathbf F_n]_{kk}=0$, and $[\boldsymbol\nu_n]_k=\bar\gamma_{k,n}\sigma_k^2/g_{kk,n}$. Since $\bar\gamma_{k,n}\le\gamma_{k,n}$ and turning off beams removes only nonnegative interference, the original SINRs give $\mathbf p_n\succeq\mathbf F_n\mathbf p_n+\boldsymbol\nu_n$. As $\boldsymbol\nu_n\succ\mathbf0$, scaling by $\operatorname{diag}(\mathbf p_n)$ shows that the row sums of the similar matrix are $[\mathbf F_n\mathbf p_n]_k/p_{k,n}<1$, so the spectral radius of $\mathbf F_n$ is less than one and $(\mathbf I-\mathbf F_n)^{-1}$ is entrywise nonnegative. Hence, \eqref{eq:app_linear} is solved by $\mathbf p_n'=(\mathbf I-\mathbf F_n)^{-1}\boldsymbol\nu_n\preceq\mathbf p_n$, which attains all targets exactly without increasing any beam power, so the transmit-power and power-gating constraints remain satisfied.

Applying this construction with $C_k^{\mathrm{tar}}=\min\{C_k,D_k\}\ge C_k^{\min}$ yields a feasible design of $\mathcal P_0$ that removes all oversupply while keeping the undersupplied volumes unchanged. Hence, every optimum of $\mathcal P_0$ satisfies $C_k\le D_k$, and $\delta_k=d_k-C_k/\kappa\in[0,\bar\delta_k]$ gives a feasible point of $\mathcal P_\delta$ with objective $v_0/\kappa^2$, i.e., $v_\delta\le v_0/\kappa^2$. Conversely, any feasible point of $\mathcal P_\delta$ satisfies $C_k/\kappa\ge d_k-\delta_k\ge c_k^{\min}$. Recovering its powers for $C_k^{\mathrm{tar}}=\min\{C_k,D_k\}$ gives a feasible design of $\mathcal P_0$ whose normalized objective is $\sum_k[d_k-C_k/\kappa]_+^2\le\sum_k\delta_k^2$. Thus, $v_0/\kappa^2\le v_\delta$, so $v_\delta=v_0/\kappa^2$, and the design recovered from any optimum of $\mathcal P_\delta$ is optimal for $\mathcal P_0$. 

The proof is complete.

\section{Proof of Lemma \ref{lem:curvature}}
\label{app:curvature}
\renewcommand{\theequation}{B.\arabic{equation}}
\setcounter{equation}{0}
For a fixed slot $n$ and illuminated-cell pair $(k,i)$, write $\mathbf c=\mathbf c_{ki,n}$, $\mathbf a=\mathbf a_{k,n}$, and $y=y_{ki,n}$. The corresponding contribution to $F_Z$ is $\phi(\mathbf Z_n)=\frac12|s(\mathbf Z_n)-y|^2$, where $s(\mathbf Z_n)=\sum_mc_me^{j\mathbf a^{\mathrm T}\mathbf z_{m,n}}$. Vectorizing the real positions gives $\|\nabla s\|_2^2=\|\mathbf a\|_2^2\|\mathbf c\|_2^2$, $\|\nabla^2s\|_2\le\|\mathbf a\|_2^2\|\mathbf c\|_\infty$, and $|s-y|\le\|\mathbf c\|_1+|y|$.

Since the position coordinates are real, $\nabla\phi=\Re\{(s-y)^*\nabla s\}$. Differentiating once more gives the Hessian
\begin{equation}
 \nabla^2\phi
 =\Re\!\left\{\nabla s(\nabla s)^{\mathrm H}+(s-y)^*\nabla^2s\right\}.
 \label{eq:hessidentity}
\end{equation}
Taking spectral norms in \eqref{eq:hessidentity}, using $\|\Re\{\mathbf A\}\|_2\le\|\mathbf A\|_2$ and $\|\nabla s(\nabla s)^{\mathrm H}\|_2=\|\nabla s\|_2^2$, yields
\begin{equation}
 \|\nabla^2\phi\|_2
 \le\|\nabla s\|_2^2+|s-y|\,\|\nabla^2s\|_2.
 \label{eq:hessbound}
\end{equation}

With $\mathbf W$ and $\mathbf Y$ fixed, these bounds hold uniformly over the positions. Summing \eqref{eq:hessbound} over the illuminated-cell pairs in slot $n$ gives $\|\nabla_{\operatorname{vec}(\mathbf Z_n)}^2F_Z\|_2\le L_n$, with $L_n$ given in \eqref{eq:Lz}.

For $\Delta\mathbf Z_n=\mathbf Z_n-\mathbf Z_n^{(s)}$, the second derivative of $F_Z$ along the segment from $\mathbf Z^{(s)}$ to $\mathbf Z$ is at most $\sum_nL_n\|\Delta\mathbf Z_n\|_F^2$, since the objective separates across slots. Taylor's theorem with integral remainder therefore yields
\begin{equation}
\begin{aligned}
 &F_Z(\mathbf Z)-F_Z(\mathbf Z^{(s)})-\sum_{n,m}(\mathbf g_{m,n}^{(s)})^{\mathrm T}
 (\mathbf z_{m,n}-\mathbf z_{m,n}^{(s)})\\
 &\le\int_0^1(1-t)\sum_nL_n\|\Delta\mathbf Z_n\|_F^2\,dt\\
 &=\frac12\sum_nL_n\|\Delta\mathbf Z_n\|_F^2.
\end{aligned}
\label{eq:app_taylor_bound}
\end{equation}
Rearranging \eqref{eq:app_taylor_bound} gives the quadratic upper bound in \eqref{eq:zmaj}. 

The proof is complete.

\bibliographystyle{IEEEtran}
\bibliography{mybib}
\end{document}